\pdfoutput=1
\begingroup\expandafter\expandafter\expandafter\endgroup
\expandafter\ifx\csname pdfsuppresswarningpagegroup\endcsname\relax
\else
\fi

\documentclass[final]{elsarticle}

\usepackage[utf8]{inputenc}

\usepackage[hyphens]{url}
\biboptions{sort&compress, square, comma}
\usepackage[breaklinks=true, linkcolor=blue, citecolor=blue, colorlinks=true]{hyperref}

\usepackage{graphicx}
\usepackage{caption}
\usepackage{subcaption}

\usepackage{longtable,booktabs,multirow}
\usepackage[version=4]{mhchem} 
\usepackage{latexsym,amsmath,amssymb}

\usepackage[T1]{fontenc}
\usepackage[english]{babel}
\usepackage{csquotes}
\usepackage{textcomp}

\usepackage{bm}

\usepackage[binary-units]{siunitx}
\DeclareSIUnit\atm{atm}

\graphicspath{{./figures/}}

\usepackage[textsize=footnotesize,textwidth=2.25cm]{todonotes}

\usepackage{algpseudocode}
\usepackage{algorithm}

\newcommand{\myVector}[1]{\ensuremath{\bm{#1}}}
\newcommand{\myMatrix}[1]{\ensuremath{\bm{#1}}}
\newcommand{\BigO}[1]{\ensuremath{O}(#1)}
\newcommand{\RU}{\ensuremath{\mathfrak{R}}}
\newcommand{\mixture}[1]{\ensuremath{\overline{#1}}}

\newcommand{\speciesk}[1]{\ensuremath{#1_k}}
\newcommand{\NReactions}{\ensuremath{N_{\mathrm{reac}}}}
\newcommand{\NSpecies}{\ensuremath{N_{\mathrm{sp}}}}

\def\CC{{C\nolinebreak[4]\hspace{-.05em}\raisebox{.4ex}{\footnotesize ++}}}

\journal{Computer Physics Communications}
\makeatletter
\def\ps@pprintTitle{%
  \let\@mkboth\@gobbletwo
  \renewcommand{\@oddhead}{}%
  \renewcommand{\@evenhead}{}%
  \renewcommand{\@evenfoot}{}%
  \renewcommand{\@oddfoot}{}%
}
\makeatother

\begin{document}

\begin{frontmatter}



\title{Accelerating finite-rate chemical kinetics with coprocessors: comparing vectorization methods on GPUs, MICs, and CPUs}


\author[cse]{Christopher P.~Stone\corref{cor1}}
\ead{chris.stone@computational-science.com}
\author[osu]{Andrew T.~Alferman}
\author[osu]{Kyle E.~Niemeyer}
\ead{kyle.niemeyer@oregonstate.edu}
\ead[url]{https://niemeyer-research-group.github.io}

\cortext[cor1]{Corresponding author}

\address[cse]{Computational Science and Engineering, LLC \\
              Athens, GA 30605, USA}
\address[osu]{School of Mechanical, Industrial, and Manufacturing Engineering\\
              Oregon State University, Corvallis, OR 97331, USA}

\begin{abstract}
Accurate and efficient methods for solving stiff ordinary differential equations (ODEs) are a critical component of turbulent combustion simulations with finite-rate chemistry.
The ODEs governing the chemical kinetics at each mesh point are decoupled by operator-splitting allowing each to be solved concurrently.
An efficient ODE solver must then take into account the available thread and instruction-level parallelism of the underlying hardware, especially on many-core coprocessors, as well as the numerical efficiency.
A stiff Rosenbrock and a nonstiff Runge--Kutta ODE solver are both implemented using the single instruction, multiple thread (SIMT) and single instruction, multiple data (SIMD) paradigms within OpenCL.
Both methods solve multiple ODEs concurrently within the same instruction stream.
The performance of these parallel implementations was measured on three chemical kinetic models of increasing size across several multicore and many-core platforms. Two separate benchmarks were conducted to clearly determine any performance advantage offered by either method.
The first measured the run-time of evaluating the right-hand-side source terms in parallel and the second benchmark integrated a series of constant-pressure, homogeneous reactors using the Rosenbrock and Runge--Kutta solvers.
The right-hand-side evaluations with SIMD parallelism on the host multicore Xeon CPU and many-core Xeon Phi co-processor performed approximately three times faster than the baseline multithreaded \CC{} code.
The SIMT parallel model on the host and Phi was \SI{13}{\percent} to \SI{35}{\percent} slower than the baseline while the SIMT model on the NVIDIA Kepler GPU provided approximately the same performance as the SIMD model on the Phi.
The runtimes for both ODE solvers decreased significantly with the SIMD implementations on the host CPU (\SIrange{2.5}{2.7}{$\times$}) and Xeon Phi coprocessor (\SIrange{4.7}{4.9}{$\times$}) compared to the baseline parallel code.
The SIMT implementations on the GPU ran \numrange{1.4}{1.6} times faster than the baseline multithreaded CPU code; however, this was significantly slower than the SIMD versions on the host CPU or the Xeon Phi.
The performance difference between the three platforms was attributed to thread divergence caused by the adaptive step-sizes within the ODE integrators.
Analysis showed that the wider vector width of the GPU incurs a higher level of divergence than the narrower Sandy Bridge or Xeon Phi.
The significant performance improvement provided by the SIMD parallel strategy motivates further research into more ODE solver methods that are both SIMD-friendly and computationally efficient.
\end{abstract}


\end{frontmatter}


\section{Introduction}
\label{intro}

Predicting turbulent combustion phenomena such as extinction and reignition with reactive computational fluid dynamics (CFD) simulations requires finite-rate chemical kinetics with detailed or reduced models.
However, the computational costs of using these can overwhelm the available computer resources.
High-fidelity combustion simulations with finite-rate kinetics must solve differential equations for the evolution of each species in the model in addition to the Navier--Stokes equations for momentum and energy.
Detailed chemical kinetic models consist of hundreds (or more) of chemical species with thousands of elementary reactions, leading to intractable storage and computational costs.
The computational cost is further increased by the stiffness of the ordinary differential equations governing chemical kinetics.
For example, in \ce{H2}\slash air combustion, the time scales of induction (\si{\micro\second}) and \ce{NO} formation (\si{\milli\second}) differ by a factor of 1000~\cite{Radhakrishnan_1984}.
This stiffness typically requires using implicit integration algorithms to solve the differential equations governing species evolution, the costs of which scale with the number of species cubed (in the worst case, associated with factorizing the Jacobian matrix)~\cite{Lu:2009gh}.

Operator splitting (e.g., Strang splitting) is commonly used to decouple the stiff chemical kinetics and nonstiff (or less stiff) convection-diffusion components of the conservation equations, as well as to reduce the size of the system of equations to be solved~\cite{Kim:1997vl,Knio:1999vd,Lanser:1999ut,Day:2000ek,Oran:2001ui,Singer_2006,Ren:2008kd,Green:2013hh}.
In this approach, the contribution of chemistry at each grid point is treated as an independent system of ordinary differential equations (ODEs) and integrated over the specified CFD time step.
That is, a large partial differential equation (PDE) system is broken into a sequence of smaller ODE systems, one for each grid point.
The species and the temperature equations are integrated in time using a constant-pressure or constant-volume assumption.
The CFD time-step size must be relatively small to avoid large splitting errors caused by thermal expansion, diffusion, and convection.
As a whole, solving all of the individual ODEs is far less expensive than a fully coupled PDE system.
Yet even with this simplification, the computational cost of solving finite-rate kinetics via local initial value problems can consume \SIrange{75}{99}{\percent} of the runtime in CFD simulations~\cite{Tonse:2003bm,Liang:2007dw,Shi:2010il,Cuoci:2013ef}.

Operator splitting provides a vast amount of parallelism since each ODE system can be solved concurrently.
Several recent studies~\cite{Spafford:2010ky,Niemeyer:2011uw,Shi:2011,Shi:2012,Stone:2013b,Niemeyer:2014jcp,Sewerin:2015hj,Curtis:2016} investigated using graphics processing units (GPUs) (i.e., accelerators) to solve the ODEs in parallel.
The kinetics ODEs are often solved with implicit backward differentiation formula (BDF) methods (e.g., VODE~\cite{Hindmarsh:2005}).
However, Stone and Davis~\cite{Stone:2013b} and Niemeyer and Sung~\cite{Niemeyer:2014jcp} demonstrated that even moderately stiff ODEs can be efficiently integrated using explicit Runge--Kutta (RK) methods on GPUs by solving many ODE systems in parallel.
For example, Stone and Davis reported a speedup of 23 $\times$ the baseline, single-core VODE CPU solver, while Niemeyer and Sung obtained a speedup of 57 $\times$ compared to a six-core OpenMP (CPU) VODE solver.
The impressive RK performance on the GPU holds even when the single-core VODE solve is parallelized linearly across multiple cores (e.g., 16-core platform).

GPUs achieve high throughput rates by combining wide vector processing, high memory bandwidth, and fast thread context-switching to hide memory latency.
NVIDIA CUDA-based GPUs implement the \emph{single instruction, multiple thread} (SIMT) vector processing paradigm which allows up to 32 threads to execute the same operation concurrently on each processor.
The high computational efficiency of the RK schemes reported above can be attributed to the fact that they have far fewer logical branches compared with the more elegant---and more complicated---BDF methods.
This allows explicit RK methods to make more efficient use of the vector processing capabilities of the GPU, a major source of their performance.
The high parallel efficiency of the RK methods can overcome their lower numerical efficiency under certain conditions.


Vector processing is a key performance feature of other many-core accelerator devices as well as most modern CPUs used in high-performance computing environments.
For example, Intel Xeon Phi (MIC) accelerators have 512-bit \emph{single instruction, multiple data} (SIMD) functional units within each core that can complete eight double-precision operations in parallel each cycle.
Modern CPUs with 256-bit AVX (or AVX2) SIMD units can complete four double operations concurrently, and future Intel Xeon CPUs and Xeon Phi devices are expected to have similar 512-bit capabilities.
Because of the high performance available from these HPC devices and the high cost of the ODE integration, adopting GPU-like, SIMD-friendly algorithms is desirable to achieve their full potential.

In this study, we compare two vectorization approaches for integrating the numerous ODE systems in parallel on modern multicore and many-core HPC platforms.
Before presenting the parallel implementation, we first introduce two ODE integration algorithms well suited for SIMD parallel processing, and three common chemical kinetics models that will be used.
We then present benchmark results using the models and discuss the performance using the various methods.
Finally, we summarize our study and provide conclusions and some recommendations for future research.

\section{ODE integration methods}
\label{sec:ode_solvers}

ODE solvers seek to advance a set of $N$ dependent variables \myVector{u}$(t)$ through time $t$ from an initial time $t_i$ to a final time $t_f$ through the action of the right-hand-side (RHS) function \myVector{f}(\myVector{u}). This can be expressed as
\begin{equation}
    \frac{d \myVector{u}}{d t} = \myVector{f}(\myVector{u}),
    \quad t_i \leq t \leq t_f \;.
    \label{eqn:ode}
\end{equation}
Here, we have assumed that the ODE system is autonomous, i.e., \myVector{f} is not a function of $t$.
A variety of techniques can be used to solve Eq.~\eqref{eqn:ode}, but all methods advance \myVector{u}$(t)$ to \myVector{u}$(t+h)$, where $h$ is an adjustable integration step size.

Integration algorithms are generally classified into two major categories: multistep and one-step methods.
Multistep methods use past time steps (e.g., \myVector{u}$(t-h)$, \ldots, \myVector{u}$(t-4h)$), while one-step methods start with only $\myVector{u}(t)$.
That is, one-step methods treat each integration step as a new integration problem.
Both classes of methods adapt $h$ in order to maintain the local truncation error (LTE) within a user-specified tolerance.
Multistep BDF methods such as VODE~\cite{Hindmarsh:2005} can also adjust the numerical order ($p$) of the method between time-steps to control the LTE.
BDF methods start as first- or second-order and take several time-steps to reach their maximum order.
One-step methods have a fixed order for all time steps.
For further details on the taxonomy of ODE solvers and stability conditions, we refer readers to the book on numerical methods for stiff systems of ODEs by Hairer and Wanner~\cite{Hairer:1996}.

The Runge--Kutta (RK) family of implicit and explicit methods are one-step methods widely used to solve both stiff and nonstiff ODE systems.
A generic $s$-stage RK method for advancing the system $\myVector{u}$ from time $t_n$ to $t_{n+1}$ is written as
\begin{equation}
    \myVector{u}_{n+1} = \myVector{u}_{n} + \sum_{i=1}^{s} b_i \myVector{k}_i \;, \label{eqn:rk}
\end{equation}
where
\begin{equation}
    \myVector{k}_i = h \myVector{f}\left( \myVector{u}_{n} + \sum_{j=1}^{s} a_{ij} \myVector{k}_j \right) \;,
\end{equation}
and $a_{ij}$ and $b_i$ are the constant parameters that define the algorithm.
Several types of RK methods exist, depending on the structure of the coefficient matrix $a_{ij}$.
Explicit RK (ERK) methods are obtained when $a_{ij}$ is strictly lower-triangular (i.e., $a_{ii} = 0$), fully implicit RK (FIRK) methods are obtained when $a_{ij}$ is fully populated, and singly diagonally implicit RK (SDIRK) methods are a special case obtained when $a_{ij}$ is lower triangular with $a_{ii} = \gamma$ (i.e., with a constant diagonal coefficient).


Examining Eq.~\eqref{eqn:rk}, we see that the FIRK and SDIRK methods are implicit, i.e., $\myVector{k}_i$ depends upon itself.
This characteristic results in a system of non--linear equations commonly solved with the iterative Newton--Raphson method.
The Newton iterative solver is a major expense for both implicit RK methods since they must compute (or approximate) and factorize the Jacobian matrix $\myVector{J}$ of the ODE system, i.e., $\myVector{J} \equiv \partial \myVector{f} / \partial \myVector{u}$.


ERK methods are efficient for nonstiff problems but are only conditionally stable.
As such, they are generally inefficient for stiff problems because the step-size $h$ is limited by stability and not by the desired accuracy.
ERK methods do not require the calculation of the Jacobian matrix (and the associated cost of solving the linear matrix systems) since they are fully explicit.
This reduces the storage requirements and computational costs for each step considerably compared to implicit methods.
The lower cost per-step of ERK may, at times, overcome the larger number of steps often required by explicit methods relative to implicit methods.
Implicit methods can typically take step sizes on the order of the CFD application's time step.

In this study, we used the five-stage, fourth-order accurate embedded Runge--Kutta--Fehlberg (RKF45) ERK solver.
\ref{app:rkf} contains the RKF parameters $\myMatrix{a}$, $\myVector{b}$, $\myVector{\hat{b}}$, and $\myVector{c}$.
The embedded fourth-order method ($p=4$) is solved simultaneously with the fifth-order method; the difference between these two solutions is used to estimate the LTE and adapt $h$ to meet the specified accuracy.

The Rosenbrock (ROS) family of one-step methods have much in common with RK methods.
ROS can be described as solving a linearized version of Eq.~\eqref{eqn:rk}.
This leads to the following $s$-stage ROS scheme~\cite{Hairer:1996}
\begin{align}
    \myVector{k}_i &= h \myVector{f}\left( \myVector{y}_{n} + \sum_{j=1}^{i-1} \alpha_{ij} \myVector{k}_j \right) + h \myVector{J} \sum_{j=1}^{i} \gamma_{ij} \myVector{k}_j \;, \quad i = 1, \ldots, s \\
    \myVector{y}_{n+1} &= \myVector{y}_{n} + \sum_{i=1}^{s} b_i \myVector{k}_i \;,
    \label{eqn:ros}
\end{align}
where
$\alpha_{ij}$, $\gamma_{ij}$, and $b_i$ are the unique method coefficients.
ROS methods are usually designed so that $\alpha_{ij}$ is lower triangular and each stage can be solved sequentially; in addition, $\gamma_{ii} = \gamma \; \forall \; i$.
A direct implementation of Eq.~\eqref{eqn:ros} requires at each stage $i$ the solution of a linear system with the matrix $I - h \gamma_{ij} \myVector{J}$ for $\myVector{k}_i$, which involves $N^2$ multiplications for $(\gamma_{ii} h) \myVector{J}$, as well as the matrix-vector multiplication $J \sum \gamma_{ij} \myVector{k}_j $.
Transforming Eq.~\eqref{eqn:ros} eliminates these expensive operations:
\begin{align}
    \left( \frac{1}{h \gamma_{ii}} \myVector{I} - \myVector{J} \right) \myVector{u}_i &= \myVector{f} \left( \myVector{y}_n + \sum_{j=1}^{i-1} a_{ij} \myVector{u}_j \right) + \sum_{j=1}^{i-1} \left( \frac{c_{ij}}{h} \right) \myVector{u}_j \;, \quad i = 1, \ldots, s \\
    \myVector{y}_{n+1} &= \myVector{y}_n + \sum_{i=1}^s m_j \myVector{u}_j \;,
\end{align}
where
\begin{align}
    \myVector{u}_i &= \sum_{j=1}^i \gamma_{ij} \myVector{k}_j \;, \quad i = 1, \ldots, s \;, \\
    \myVector{a} &= \bm{\alpha} \myVector{\Gamma}^{-1} \;, \\
    \myVector{c} &= \gamma \myVector{I} - \myVector{\Gamma}^{-1} \;, \\
    \myVector{m} &= \myVector{b} \myVector{\Gamma}^{-1} \;,
\end{align}
and $\myVector{\Gamma} = (\gamma_{ij})$.
We implemented the four-stage, fourth-order ROS4 scheme of Hairer and Wanner~\cite{Hairer:1996}---corresponding to their $L$-stable fourth-order Rosenbrock method, where $\gamma = 0.572816$---also available in the FATODE package~\cite{Zhang:2014}.
\ref{app:ros4} contains the ROS4 parameters $\myMatrix{a}$, $\myMatrix{c}$, $\myVector{m}$, $\alpha_i$, and $\gamma_i$ needed to reimplement the method, although Fortran implementations are provided by Hairer and Wanner~\cite{ROS4:1992,Hairer:1996} and Zhang and Sandu~\cite{FATODE:1.2,Zhang:2014}.

The major distinction between the Rosenbrock and implicit RK methods lies in the role of the Jacobian matrix.
\myVector{J} is only used to converge the nonlinear systems in the fully implicit schemes and is not part of the final solution.
As such, \myVector{J} can be approximated or reused over many steps so long as the Newton iteration converges economically.
Conversely, \myVector{J} appears explicitly in Eq.~\eqref{eqn:ros} and must be computed at each step in ROS methods.
This requirement increases the computational cost of Rosenbrock methods if the construction and factorization of the Jacobian is costly.

The non-iterative nature of the ROS methods has several advantages from a parallel processing point of view.
As discussed earlier, ERK methods perform favorably on GPUs primarily due to their simplicity and low level of divergence relative to the more complicated BDF methods.
This allows their high vector parallel efficiency to overcome their lower numerical efficiency.
Unlike ERK methods, ROS methods are L-stable and can handle stiff ODEs.
Since they do not require any iterative solution, they can be implemented efficiently in a SIMD environment much like ERK methods.
We implemented the fourth-order accurate ROS method (ROS4) to permit direct comparisons with the RKF45 method.
For further details on the Rosenbrock method used here, see Hairer and Wanner~\cite{Hairer:1996} or Zhang and Sandu~\cite{Zhang:2014}.



\section{Chemical kinetics model}
\label{sec:rate_model}

In this section, we introduce the model chemical kinetics problem that will be used throughout the performance benchmarks.

The following ODE system governs the time evolution of {\NSpecies} chemical species and energy for a constant-pressure, gas-phase combustion process at each grid point or cell:
\begin{align}
   \dot{Y}_k &= \frac{\dot{\omega}_k W_k}{\rho} \label{eqn:ydot} \text{ and} \\
   \dot{T} &= -\frac{1}{c_p}\sum_{k=1}^{\NSpecies} h_k \dot{\omega}_k \label{eqn:tdot} \;,
\end{align}
where $Y_k$, $W_k$, $h_k$, and  $\dot{\omega}_k$ are the mass fraction, molar mass, enthalpy, and molar production rate for species $k$; and $T$, $\rho$, and $c_p$ are the mixture temperature, density, and specific heat at constant pressure ($c_p = \sum_{k=1}^{\NSpecies} Y_k c_{p,k}$, where $c_{p,k}$ is the constant-pressure specific heat in mass units).
Equations~\eqref{eqn:ydot} and \eqref{eqn:tdot} are closed by the equation of state for an ideal gas, $p = \rho \mixture{R} T$, where $p$ is the thermodynamic pressure and \mixture{R} is gas constant of the mixture.
A constant-volume process, i.e., where $\rho$ is constant, can be modeled by replacing $c_p$ with $c_v$ and $h_k$ with $u_k$ in Eq.~\eqref{eqn:tdot}.

The net molar production rate terms ($\dot{\omega}$) are nonlinear functions of pressure $p$, $T$, and the species molar concentrations $[X_k]$.
They are also the source of the stiffness in the ODE system and their calculation is generally the most computationally intensive component of the integration.
They are expressed as
\begin{equation}
\dot{\omega}_k = \sum_i^{{\NReactions}_{, k}} c_i
        \left( \nu_{ki}'' - \nu_{ki}' \right)
        \left( k_{f,i} \prod_j^{{\NSpecies}_{, i}} [X_{j}]^{\nu_{ji}'} - k_{r,i} \prod_j^{{\NSpecies}_{, i}} [X_{j}]^{\nu_{ji}''} \right) \;,
\label{eqn:wdot}
\end{equation}
where ${\NReactions}_{, k}$ is the number of reactions involving species $k$, $c_i$ is a parameter accounting for any third-body and\slash or pressure effects, $\nu_{ki}^{\prime}$ and $\nu_{ki}^{\prime\prime}$ are the reactant and product stoichiometric coefficients for species $k$ in reaction $i$, ${\NSpecies}_{, i}$ is the number of reactants and products in reaction $i$, and $k_f$ and $k_r$ are the forward and reverse reaction rate coefficients.
Details regarding the third-body and pressure fall-off effects embodied in $c_i$ are given by Niemeyer et al.~\cite{Niemeyer:2016aa}.

The {\NReactions} forward rate constants are given in Arrhenius form as
\begin{equation}
k_{f,i} = A_i\, T^{\beta_i}\, exp \left( -\frac{E_i}{\RU T} \right) \;,
\label{eqn:kfor}
\end{equation}
where $\RU$ is the universal gas constant.
If reaction $i$ is irreversible, $k_{r,i}$ is zero.
Explicit reverse Arrhenius rate coefficients can be given.
Otherwise, they are computed as a function of the equilibrium constant $K_{c,i}$
\begin{equation}
k_{r,i} = \frac{k_{f,i}}{K_{c,i}} \label{eqn:krev} \;,
\end{equation}
where $K_{c,i}$ is computed as
\begin{align}
K_{c,i} &=  \left( \frac{p_0}{\RU T} \right)^{ \sum_{j}^{{\NSpecies}_i} \nu_{ji} } K_{p,i} \\
K_{p,i} &= \exp\left[ \sum_j^{{\NSpecies}_i} \left( \nu_{ji}'' - \nu_{ji}' \right) \left( \frac{S_j}{\RU} - \frac{H_j}{\RU T} \right) \right] \;,
\label{eqn:kc}
\end{align}
where $p_0$ is the standard pressure at one atmosphere (in the appropriate units), and $S_j$ and $H_j$ are the standard-state entropy and enthalpy of species $j$ in molar units.
Temperature-dependent thermodynamic properties (e.g., $c_{p,j}$, $H_j$, $S_j$) are computed from polynomial fits using the following formulas:


\begin{align}
  \frac{c_{p,j}}{\speciesk{R}} &= \sum_{k=1}^{4} A_{k} T^{k-1} \nonumber \\
  \frac{H_j}{\RU} &= \sum_{k=1}^{5} A_{k} T^{k} + \frac{A_{6}}{T} \label{eqn:tfits} \\
  \frac{S_j}{\RU} &= A_{1}\, \log(T) + \sum_{k=2}^{5} A_{k} T^{k-1} - \frac{A_{6}}{T} + A_{7} \nonumber \;.
\end{align}

\noindent The polynomial coefficients $A_{k}$ are taken from the NASA seven-term polynomial~\cite{Mcbride:1993} database.
Two or more sets of coefficients are typically used, with each valid over a specified temperature range.







\section{Parallel integrator implementations}
\label{sec:parallel_implementations}


As noted earlier, we wish to solve thousands of ODE systems concurrently on multicore and many-core devices.
Before presenting the parallel implementation strategies, it is necessary to define terminology that spans the various HPC architectures.
These devices offer two distinct levels of parallelism: multiple processing elements (e.g., multiple cores) and SIMD vector processing within each processing element.
Vector instructions are issued by the processing elements and executed in SIMD parallel fashion across multiple data streams.
In this paradigm, logical flow is controlled at the processing-element level and fine-grain data parallelism is implemented within each processing element.
For this discussion a lane represents a single slot within the SIMD unit, logical threads issue vector instructions, one or more threads occupy a single processing element (e.g., \emph{hyperthreading}), and threads may execute separate logical tasks.



The numbers of species ({\NSpecies}) in the chemical kinetic models of interest range between {\BigO{10}} and {\BigO{100}} and the number of reactions ({\NReactions}) scales linearly (i.e., ${\NReactions} \approx 5 {\NSpecies}$~\cite{Lu:2009gh}).
The number of independent ODE systems ($N_{\text{ode}}$) range from \BigO{\num{e4}}--\BigO{\num{e6}} \emph{per device}~\cite{Unat:2015:IJHPCA} in 3-D combustion simulations.
Two SIMD parallel processing strategies can be designed based on these expected values.

In the first approach, a single thread solves multiple ODE systems together by mapping each ODE to a separate SIMD lane.
For example, each lane evaluates Eqs.~\eqref{eqn:ydot}--\eqref{eqn:tfits} with a unique set of $y_k$ and $T$ values at a given time.
Multiple sets of ODEs can be integrated by separate threads on other PEs.
Since vectorization is applied across only the breadth of the set of ODEs, we refer to this approach as \emph{shallow} vectorization.
This technique is analogous to the GPU\slash CUDA-specific \emph{per-thread} approach demonstrated by Stone and Davis~\cite{Stone:2013b} and Niemeyer and Sung~\cite{Niemeyer:2014jcp}.

Conversely, in the second approach, a thread integrates a single ODE system but evaluates Eqs.~\eqref{eqn:ydot}--\eqref{eqn:tfits} using the data parallel vector operations.
For example, all {\NReactions} forward rate coefficients given by Eq.~\eqref{eqn:kfor} can be evaluated in parallel.
More complex terms such as Eq.~\eqref{eqn:wdot} require parallel reductions for each species $k$.
In this approach, vectorization is applied through the depth of each ODE system.
This method is therefore termed \emph{deep} vectorization.
This technique is analogous to the \emph{per-block} (or \emph{per-thread-block}) CUDA-specific method demonstrated by Stone and Davis~\cite{Stone:2013b}.
As before, multiple ODEs are still evaluated concurrently across all of the available processing elements.
For example, each OpenMP thread would solve a separate ODE system.

Stone and Davis~\cite{Stone:2013b} demonstrated these two vectorization techniques for a single chemical kinetic model \cite{lu:2005,Lignell:2007} with 19 species, 167 elemental reactions, and ten quasi-steady-state intermediate species on an NVIDIA C2050 (Fermi) GPU.
They used CUDA implementations of the RKF45 and VODE BDF solvers to integrate hundreds of thousands of stiff ODE systems.
They reported a \SI{20.2}{$\times$} speedup for CUDA-RKF45 but only \SI{7.7}{$\times$} with CUDA-VODE relative to the single CPU core VODE (CPU-VODE) runtime with the shallow vectorization approach.
When normalized by the RKF45 solver on the CPU, the CUDA-RKF45 was \SI{28.6}{$\times$} faster with shallow vectorization.
With the deep vectorization method, they reported speedups of only \SI{10.7}{$\times$} and \SI{7.3}{$\times$} with CUDA-RKF45 and CUDA-VODE, respectively, relative to CPU-VODE.
They attributed the lower speedup of both solvers with deep vectorization to the small model, e.g., 19 species is much smaller than the effective SIMD width (32 lanes) on the CUDA device.
They also attributed the superior vector efficiency of the RKF45 solver to its simplicity relative to the VODE.
Niemeyer and Sung~\cite{Niemeyer:2014jcp} similarly reported a speedup of \SI{57}{$\times$} using a CUDA GPU with shallow vectorization compared with execution on a six-core CPU.
They used a stabilized Runge--Kutta--Chebyshev (RKC) ODE solver with a moderately stiff chemical kinetic model with 53 species and 634 irreversible reactions.

Both of these prior studies used CUDA GPUs.
CUDA offers a particularly straightforward approach to implement the shallow vectorization paradigm.
In the CUDA development environment, explicit SIMD instructions are not necessary.
Instead, the runtime gathers sets of CUDA threads into \emph{warps} and maps these warps to individual streaming multiprocessors.
Threads within the same warp all execute the same instruction in lockstep following the SIMT paradigm.
That is, CUDA threads map to lanes within the individual streaming-multiprocessor vector units.
This effectively permits a serial implementation to be replicated across all vector lanes and all processors.
As Stone and Davis~\cite{Stone:2013b} noted, implementing deep vectorization is much more complicated and requires explicit synchronization and communication among CUDA threads within the same \emph{thread-block}, i.e., a collaborating team of warps.

This paradigm is largely inverted on modern multicore CPU environments including the Intel Xeon Phi, where a single CPU thread occupies the entire processing element.
The CPU thread issues explicit SIMD instructions to enact operations across the parallel lanes of the vector units.
Kroshko and Spiteri~\cite{Kroshko:2013:JOCS} demonstrated this approach in their SIMD implementation of a RODAS Rosenbrock solver. There, they reported a speed-up of \SI{1.89}{$\times$} (i.e., \SI{94}{\percent} parallel efficiency) when solving multiple systems of stiff IVPs on a cell broadband engine.

Explicit SIMD programs have historically been platform-dependent and their implementation has been quite difficult.
Instead, most developers rely upon vectorizing compilers to identify parallel loops and automatically generate SIMD code for each target platform.
Due to the structure of the ODE solver implementations and libraries, this approach generally results in an application following the deep vectorization paradigm.
That is, each ODE system is solved by a single CPU thread and the compiler vectorizes loops with fixed (i.e., known) length.
For example, the loop evaluating the forward reaction rate coefficients (Eq.~\eqref{eqn:kfor}) for all {\NReactions} can be vectorized.
Implementing shallow vectorization on the Xeon Phi or the host CPU cores requires explicit SIMD programming, a more challenging parallel programming style.
As demonstrated in the above citations, this approach appears to hold much promise for HPC platforms and may warrant the added implementation complexity.

For this study, the RKF45 and ROS4 ODE solvers were implemented using shallow vectorization with the OpenCL~\cite{Stone:2010:OPP:622179.1803953} language.
OpenCL provides SIMD datatypes of varying lengths, e.g., two to sixteen doubles per SIMD superword, as well as traditional scalar datatypes.
Scalar datatypes are suitable when relying upon compiler-generated vectorization or for SIMT (GPU) environments.
OpenCL is also platform independent which allows performance studies across multiple platforms.

We did not explicitly implement deep vectorization in OpenCL.
Instead, deep vectorization was realized through \emph{guided} compiler vectorization of the original C\slash \CC{} implementation.
Compiler vectorization directives were added to species and reaction loops within the chemical kinetics RHS function and matrix factorization and other loops within the ODE solvers to facilitate vectorization where necessary and appropriate.

As noted in Section~\ref{sec:ode_solvers}, the RK and ROS algorithms are quite similar and largely share the same logical flow.
Algorithm~\ref{alg:scalar_ode_solver} represents both solvers using traditional scalar datatypes.
The function \texttt{OneStep} advances $\myVector{u}$ from $t$ to $t+h$ using Eq.~\eqref{eqn:rk} or \eqref{eqn:ros} for RK or ROS solvers, respectively.
This function returns a trial solution $\myVector{u^*}(t+h)$ and an approximation of the LTE.
\texttt{AdjustStepSize} implements step-size size adaption based on the LTE approximation.

In contrast, Algorithm~\ref{alg:simd_ode_solver} shows the equivalent SIMD implementation.
Several SIMD functions are introduced there and their meanings are:
\begin{description}
  \item[Gather] Read multiple scalar values from arbitrary locations into a single SIMD word.
  \item[Scatter] Write the SIMD vector elements to arbitrary scalar locations.
  \item[Broadcast] Replicate a scalar value across all SIMD vector elements or lanes.
  \item[Select(mask,a,b)] Merge the SIMD words \myVector{a} and \myVector{b} based on the SIMD logical \myVector{mask}. That is, for each \emph{lane} k within the SIMD word, return a[k] if mask[k] evaluates True; otherwise, return b[k].
  \item[isLess] Logically evaluate if the vector elements are less than a given value and return a logical SIMD mask.
  \item[isGreaterEqual] Logically evaluate if the vector elements are greater than or equal to a given scalar value and return a logical SIMD mask.
  \item[Any] Return True if any vector elements evaluate True; False otherwise. This is a SIMD \emph{reduction} operation.
  \item[All] Return True if all vector elements evaluate True; False otherwise.
\end{description}
The SIMD implementation in Algorithm~\ref{alg:simd_ode_solver} is equally valid for scalar datatypes (i.e., a SIMD word width of one is a scalar datatype), and the scalar and SIMD results are numerically equivalent (to within double datatype precision).

While complex and computationally intensive, the chemical kinetics rate calculations (Eq.~\eqref{eqn:ydot}--\eqref{eqn:kc}) require no SIMD reduction operations or collective logic tests.
The exception is Eq.~\eqref{eqn:tfits} since the temperature of each SIMD vector element (i.e., a \emph{lane}) could lie within different polynomial fit ranges.
That is, different polynomial coefficients are needed for different lanes.
For two temperature ranges, we compute both polynomial equations for the desired thermodynamic quantity (e.g., $C_p$).
A SIMD masked \textbf{Select} operation then selects the correct polynomial fit depending upon the lane mask.
The number of valid temperature ranges is arbitrary~\cite{Mcbride:1993}, though two is common in practice.
More than two temperature ranges would require a more complex SIMD algorithm.

\begin{algorithm}
\caption{Scalar RK and ROS solver algorithms to integrate $N_{\text{ode}}$ independent ODE systems: Advance the initial state $y_0$ from $t_i$ to $t_f$ using step-size adaption to control the leading truncation error (LTE).}
\label{alg:scalar_ode_solver}
\begin{algorithmic}[1]
\For {$k \gets 1, N_{\text{ode}}$}
    \State $\myVector{u}(t) \gets \myVector{y}_0[k]$
    \State $t \gets t_i$
    \State $h \gets \texttt{EstimateH0}(\myVector{u}(t), \texttt{MaxError})$
    \While {$t < t_{i}$}
       \State $\myVector{u^*}(t+h), err \gets \texttt{OneStep}(\textbf{u}(t))$
      \If {$err < \texttt{MaxError}$}
          \State $\myVector{u}(t+h) \gets \myVector{u^*}(t+h)$
         \State $t \gets t+h$
      \EndIf
       \State $h \gets \texttt{AdjustStepSize}(err, \; h)$
    \EndWhile
    \State $\myVector{y}[k] \gets \myVector{u}$
\EndFor
\end{algorithmic}
\end{algorithm}

\begin{algorithm}
\caption{SIMD RK and ROS solver algorithms to integrate $N_{\text{ode}}$ ODE systems: All values are SIMD datatypes with \emph{W} vector elements per word.}
\label{alg:simd_ode_solver}
\begin{algorithmic}[1]
\For {$k \gets 1, N_{\text{ode}}$ in increments of $W$}
    \State $\myVector{u}(t) \gets \textbf{Gather}(\myVector{y}_0[k:k+W])$
    \State $t \gets \textbf{Broadcast}(t_i)$
    \State $h \gets \texttt{EstimateH0}(\myVector{u}(t), \texttt{MaxError})$
    \Repeat
       \State $\myVector{u^*}(t+h), err \gets \texttt{OneStep}(\myVector{u}(t))$
       \State $mask \gets \textbf{isLess}(err, \texttt{MaxError})$
       \If {$\textbf{Any}(mask)$}
         \State $\myVector{u}(t+h) \gets \textbf{Select}(mask, \myVector{u^*}(t+h), \myVector{u}(t))$
          \State $t \gets \textbf{Select}(mask, t+h, t)$
       \EndIf
      \State $h \gets \texttt{AdjustStepSize}(err,h)$
       \State $finished \gets \textbf{isGreaterOrEqual}(t, t_f)$
    \Until {$\textbf{All}(finished)$}
    \State $\myVector{y}[k:k+W] \gets \textbf{Scatter}(\myVector{u})$
\EndFor
\end{algorithmic}
\end{algorithm}

A major distinction of the SIMD solver implementation is that the time-step iteration loop continues for all ODE systems grouped into the same SIMD data stream until all reach $t_f$.
This can lead to inefficiency if the number of iterations varies significantly.
Both Stone and Davis~\cite{Stone:2013b} and Niemeyer and Sung~\cite{Niemeyer:2014jcp} found that this phenomena can significantly degrade performance if the initial states of the ODE systems within the same SIMD stream differ widely.

Both Algorithms~\ref{alg:scalar_ode_solver} and \ref{alg:simd_ode_solver} were implemented using OpenCL scalar and SIMD datatypes, respectively.
All computations were performed exclusively in double precision.
OpenCL supports SIMD datatypes with \numlist{2;4;8;16} vector elements per word.
This permits between 2 and 16 ODEs to be solved concurrently within each invocation of the ODE solvers.
All SIMD functions discussed above are provided suitable gather\slash scatter operations.
The scalar and SIMD algorithms were implemented separately despite their similarity due to incompatibility between the OpenCL scalar and SIMD functions and the lack of operator-overloading features.\footnote{At the time of this study, the OpenCL standard (1.2) supported SIMD vector loads\slash stores to contiguous memory locations but not strided variants.
In addition, OpenCL did not support \CC{} within device code, which precluded operator-overloading or template functions.}

\section{Results}

We performed a series of benchmarks across three fundamentally different platforms and three different chemical kinetic models to assess the performance of the ODE solvers and the chemical kinetics rate evaluations within the SIMD context.

The platforms include a NVIDIA Kepler K20m GPU, an Intel Xeon Phi SE10P (MIC) coprocessor with 61 cores, and an Intel Sandy Bridge E5-2680 CPU with eight cores and two CPUs per compute node (for 16 total cores).
The MIC and CPU OpenCL routines were compiled using the Intel OpenCL 1.2 SDK (v1.2.0.76921); Kepler routines used the NVIDIA OpenCL 1.1 driver (v331.67).
The host driver application was built using the Intel \CC{} compiler (v14.0.1).
Multithreaded (OpenMP) \CC{} implementations of the ODE solvers and RHS function evaluations were also compiled with the Intel \CC{} compiler for performance comparison.

Benchmarks on the MIC and GPU accelerators do not include communication time.
The focus of this study is computational throughput on these devices and on the host devices, not specifically the acceleration over the host offered by these accelerators.
All GPU benchmarks used 512 threads and 32 thread blocks per SMX.
This gave the highest performance on the GPU for the three kinetic models.
Also, all data is stored in \emph{global} GPU memory; no \emph{shared} or \emph{constant} memory was used for these benchmarks due to their size restrictions.

\begin{table}[htbp]
  \centering
  \caption{Details of the chemical kinetic models considered here.}
  \label{tab:mechs}
  \begin{tabular}{@{}l c c c c@{}}
    \toprule
    Name & Fuel & {\NSpecies} & {\NReactions} & Reference \\
    \midrule
    \ce{H2}\slash\ce{CO} & \ce{H2}   & 14  & 38  & \cite{Davis:2005} \\
    GRI Mech 3.0         & \ce{CH4}  & 53  & 325 & \cite{grimech3} \\
    USC Mech II          & \ce{C2H4} & 111 & 784 &  \cite{Wang:2007} \\
    \bottomrule
  \end{tabular}
\end{table}

Table~\ref{tab:mechs} shows details of the three chemical kinetic models considered in this study: the \ce{H2}\slash\ce{CO} model of Davis et al.~\cite{Davis:2005}, GRI Mech 3.0~\cite{grimech3}, and USC Mech Version II~\cite{Wang:2007}.
Most reactions are reversible and all three models contain both third-body and pressure-dependent reactions; specific details can be found in their associated references.
As noted earlier, all thermodynamic polynomial curve fits (Eq.~\eqref{eqn:tfits}) for these models use two temperature ranges.

The chemical kinetic models were interpreted using the \texttt{create\_rate\_subs} software~\cite{create-rate-subs:v1}, and the necessary species and reaction rate information saved to a binary database in turn read by the source term functions.
Jacobian matrices (needed for the ROS integrator) are constructed using first-order forward finite differences, following the approach used in CVODE~\cite{Hindmarsh:2005}.
Evaluating each Jacobian thus requires $\NSpecies+1$ RHS function evaluations, in addition to one RHS evaluation per stage.

\subsection{Performance of RHS evaluation}

\begin{figure}[htbp]
  \centering
  \includegraphics[width=0.9\textwidth]{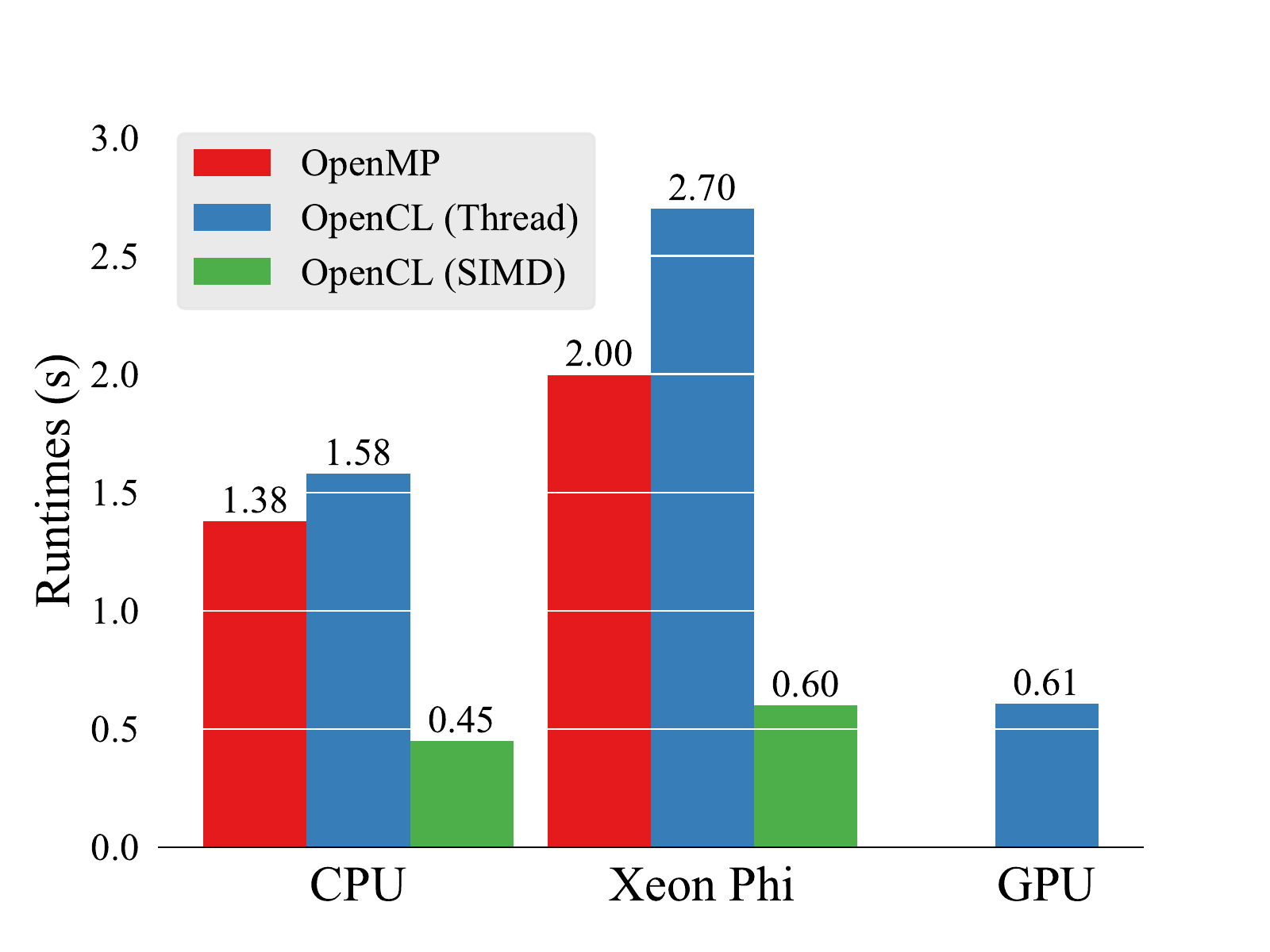}
  \caption{Wall-clock runtimes (seconds) for one million RHS evaluations of the GRI Mech 3.0 model on the host CPU, MIC, and Kepler GPU using the OpenCL SIMD (CL-SIMD) and thread-parallel (CL-Thread) implementations. OpenMP uses sixteen threads with compiler auto-vectorization.
  Data, plotting scripts, and figure file are available~\cite{figures}.}
  \label{fig:rhs_GRI Mech 3.0_best}
\end{figure}

The first set of benchmarks studied the throughput of RHS function evaluations.
Figure~\ref{fig:rhs_GRI Mech 3.0_best} shows the average runtimes for one million RHS evaluations with the GRI Mech 3.0 model on the host CPU, MIC coprocessor, and the Kepler GPU using the SIMD and thread-parallel OpenCL implementations.
These benchmarks show the best performing configurations for each device.
For the SIMD host and MIC benchmarks, the most efficient word length was twice the native size, i.e., eight-wide on the host.

The unique thermochemical input states for each RHS evaluation were generated by setting a uniform composition for all species but linearly varying the temperature over \SIrange{500}{1500}{\kelvin} to cross the polynomial temperature ranges.
This requires evaluation of all temperature--dependent branch statements, such as the polynomial curve fits for specific heat, which reduces the SIMD (or SIMT) parallel efficiency.
Even with this forced divergence, the explicit SIMD implementations improve the runtime by a factor of 3.1 on the host CPU and 3.3 on the MIC.
The OpenCL thread-parallel runtimes are considerably slower than with the OpenCL SIMD method and are slower than the OpenMP baseline.
The OpenMP baseline on the host used 16 threads and 240 threads\footnote{Each MIC core can support four hardware threads, and one core is reserved for the MIC operating system.} on the MIC.
The thread-parallel implementation on the Kepler GPU gives favorable performance and is 2.3 times faster than the baseline host runtime.
This indicates that the thread-parallel approach on the SIMD platforms (i.e., the host CPU and MIC) is inefficient for this type of application.
However, the thread-parallel SIMT method is most efficient on the GPU.

\begin{figure}[htbp]
  \centering
  \includegraphics[width=0.9\textwidth]{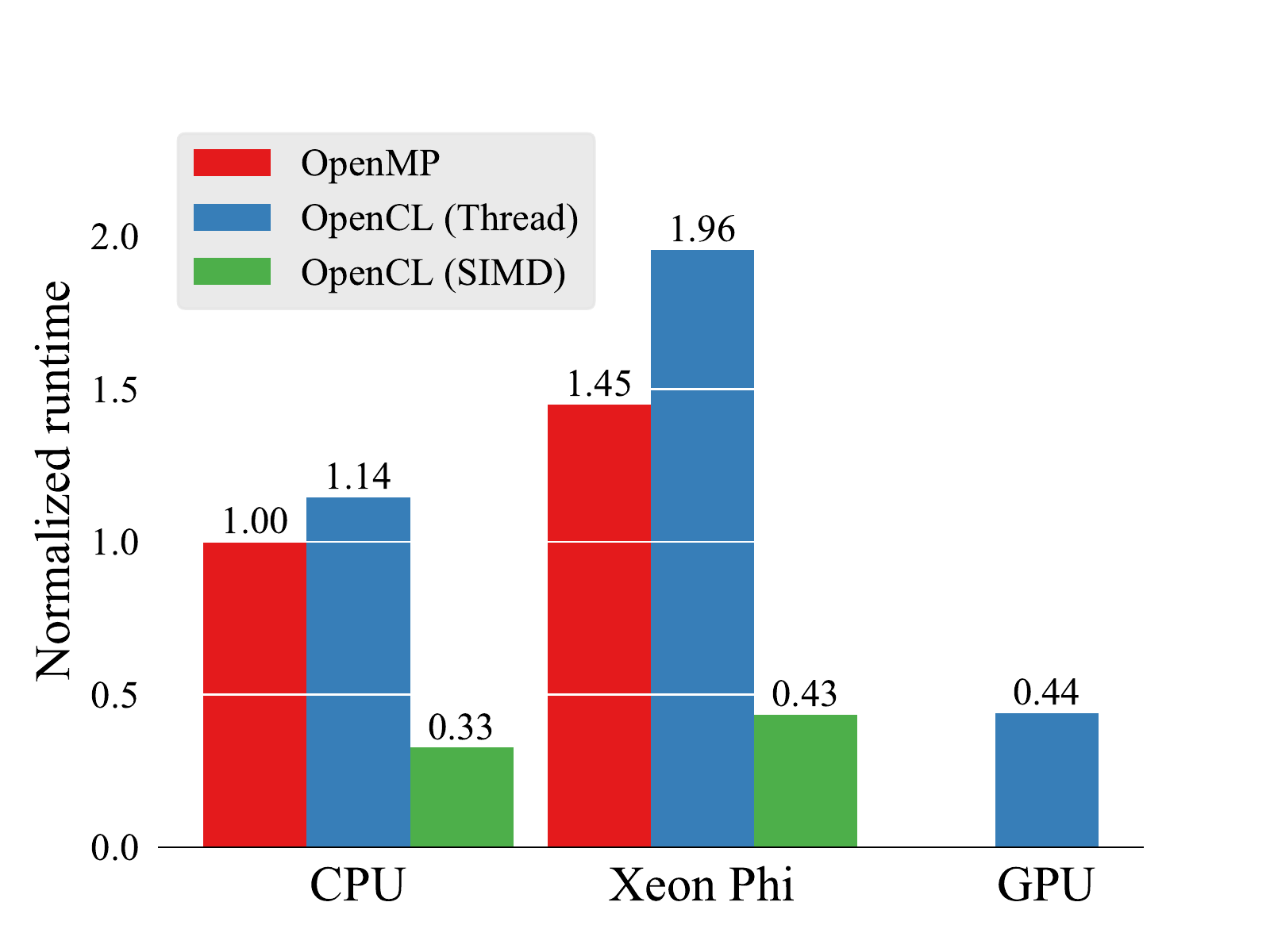}
  \caption{Normalized runtimes (speedup) for one million RHS evaluations of the GRI Mech 3.0 model on the host CPU, MIC, and Kepler GPU using the OpenCL SIMD (CL-SIMD) and thread-parallel (CL-Thread) implementations. Runtimes are normalized by the OpenMP runtime on the host CPU using compiler auto-vectorization.
  Data, plotting scripts, and figure file are available~\cite{figures}.}
  \label{fig:rhs_GRI Mech 3.0_best_speedup}
\end{figure}

Figure~\ref{fig:rhs_GRI Mech 3.0_best_speedup} shows the runtimes (from Figure~\ref{fig:rhs_GRI Mech 3.0_best}) normalized by the host runtime with OpenMP and auto-vectorization.
This gives the relative performance compared to the host baseline.
The GPU and MIC (with CL-SIMD) both give nearly identical speedup (\SI{2.3}{$\times$}) over the baseline.
These accelerators perform well compared with the baseline; however, the significant host improvement from CL-SIMD means that the host is still faster by nearly \SI{25}{\percent}.

\begin{figure}[htbp]
  \centering
  \includegraphics[width=0.9\textwidth]{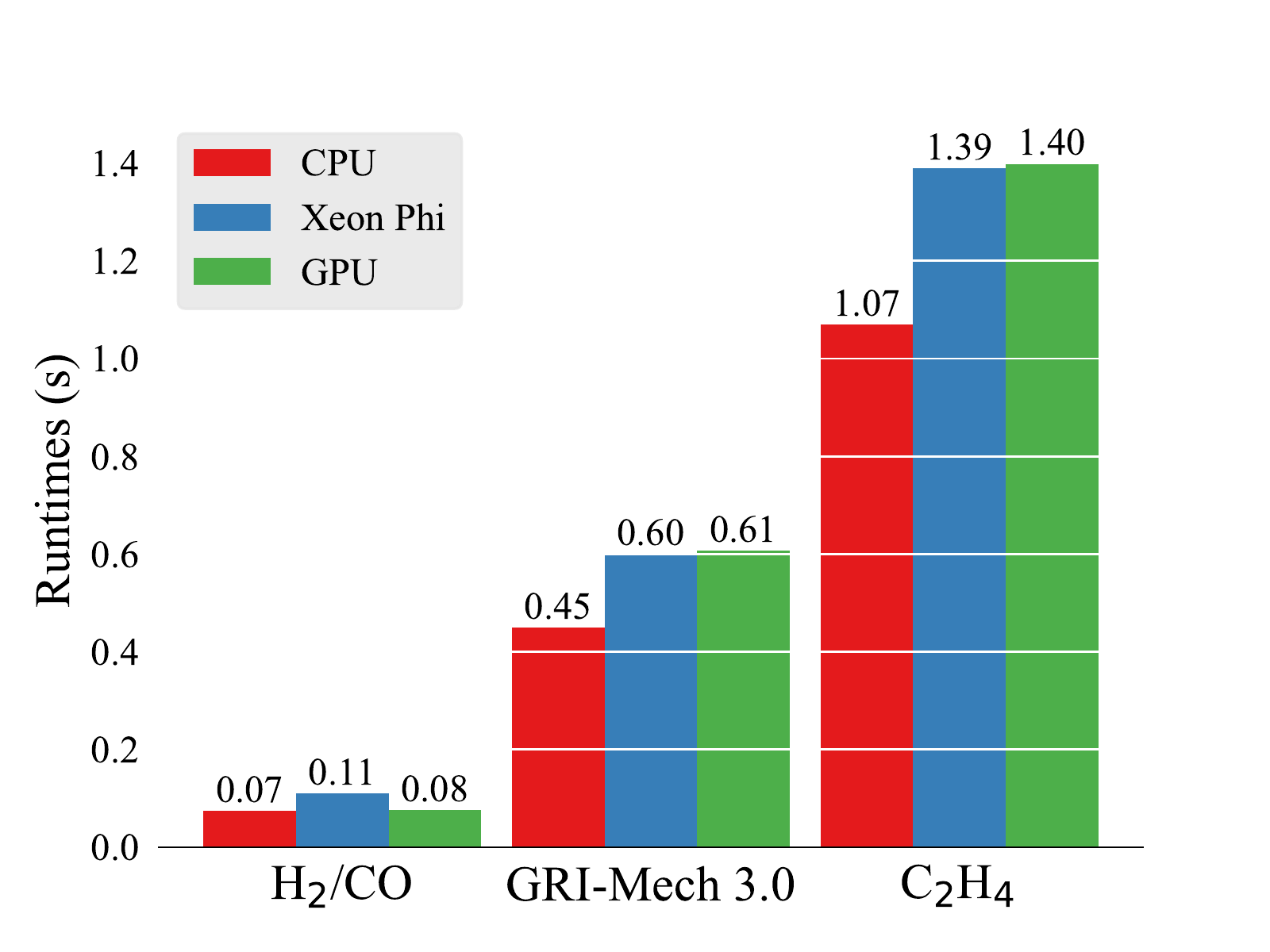}
  \caption{Wall-clock runtimes (seconds) for one million RHS evaluations for all three models in Table \ref{tab:mechs} using the SIMD method (CL-SIMD) for the CPU and MIC and the thread-parallel method (CL-Thread) for the GPU.
  Data, plotting scripts, and figure file are available~\cite{figures}.}
  \label{fig:rhs_all_best}
\end{figure}

Figure~\ref{fig:rhs_all_best} shows RHS runtimes for all three models studied using the SIMD method for the CPU and MIC and the thread-parallel method for the GPU (i.e., the fastest method for each device).
For the larger two models, the relative performance between the host and the accelerators is nearly the same.
However, the GPU and host perform equivalently for the smaller \ce{H2}\slash\ce{CO} model.




These results show that the SIMD method is quite effective for evaluating the RHS function.
The RHS function is complex; however, as noted earlier, only the thermodynamic polynomials are able to diverge across SIMD lanes.

\subsection{Performance of ODE time integration}

Let us now shift to applying this approach to time integration of the ODE systems described above.
The kinetic reaction rate evaluation function is the primary computational expense since it is used both for the RHS function and for generating the system Jacobian with finite-differences.
Unlike the previous experiment, the potential for lane divergence increases when mapping an ODE system to each lane.

The RKF45 and ROS4 solvers are single-step methods with a fixed cost per-step.
The number of steps needed to solve the ODE system may vary between systems, which causes divergence of severity depending highly on the problem.

\begin{figure}[htbp]
  \centering
  \includegraphics[width=0.9\textwidth]{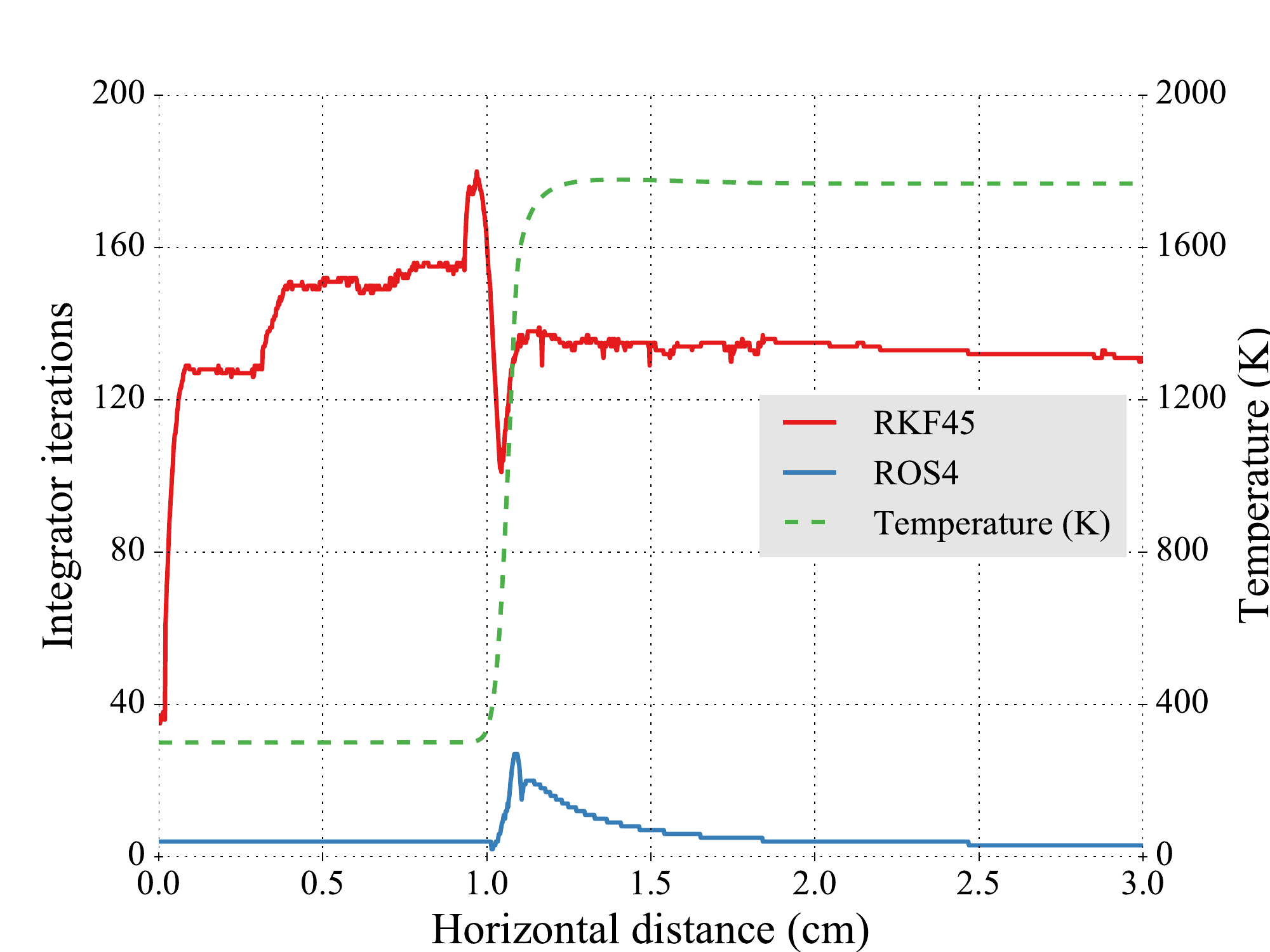}
  \caption{Temperature profile normal to the flame and the number of integrator steps needed for RKF45 and ROS4 ODE solvers.
  Data, plotting scripts, and figure file are available~\cite{figures}.}
  \label{fig:flame_profile}
\end{figure}

The parallel ODE solvers were applied to state data obtained from a stationary
one-dimensional premixed methane\slash air flame simulation.
The initial flame profile was computed using the GRI Mech 3.0 model~\cite{grimech3} and
Cantera's \texttt{FreeFlame}, a simulation tool for modeling freely propagating flat flames~\cite{Cantera},
and then interpolated onto a uniform mesh with 1601 points.
The unburned temperature at the left boundary is \SI{300}{\kelvin} and the equivalence ratio of the fresh reactants is 0.67.
The resulting state data are available openly~\cite{profiledata}.
This spatial resolution (i.e., approximately 25 mesh points across the thermal flame thickness~\cite{Aspdena:2015:PCI}) compares with that needed for a direct numerical simulation of complex phenomena such as flame-turbulence interaction.
To mimic an operator-splitting framework, the ODE systems at each mesh point are integrated independently over a fixed time of \SI{1}{\micro\second}.
This time interval represents a feasible convective time-step size for semi-implicit~\cite{Cuoci:2013ef} CFD methods.

Figure~\ref{fig:flame_profile} shows the temperature profile normal to the flame, where the thin reaction zone is evident.
The number of \emph{attempted} integrator steps (i.e., accepted and rejected steps) needed for the RKF45 and ROS4 ODE solvers are shown as well.
Both ODE solvers use the same absolute (\num{e-8}) and relative (\num{e-11}) tolerances, the same initial time-step size ($h_0$) estimation, and the same $h$-adaption algorithm.
The differences in stability characteristics of the two ODE solvers cause the differences in number of steps.
The L-stable ROS4 solver can quickly solve the largely non-reactive zones upstream and downstream of the flame with only a minimal, and largely constant, number of integrator steps.
Only in the thin flame itself does the solver need more than this minimum.
The number of steps needed by RKF45 fluctuates but is, in general, higher regardless of local conditions.
This results from the stiff conditions of the kinetics problem.
In this scenario, stability limits $h$, rather than local error as in the case of the stiff ROS4 solver.
The cost per step is not equal between RKF45 and ROS4 since the latter must also construct and factorize the Jacobian matrix.
For the GRI Mech 3.0 model, ROS4 requires \SI{10}{$\times$} more RHS function evaluations per step.
The RHS ratio is a good estimation of the overall per-step cost ratio of the two methods.

To mimic the cost of a multidimensional reactive CFD simulation, the one-dimensional domain shown in Figure~\ref{fig:flame_profile} is replicated \numlist{10;40;100;200} times vertically to give approximately \numlist{16e3;48e3;160e3;320e3} points.
These sizes were selected to approximate mesh sizes that may be encountered on a per-core (e.g., \num{16e3}) and per-device (e.g., \num{320e3}) basis.

\begin{figure}[htbp]
  \centering
  \includegraphics[width=0.9\textwidth]{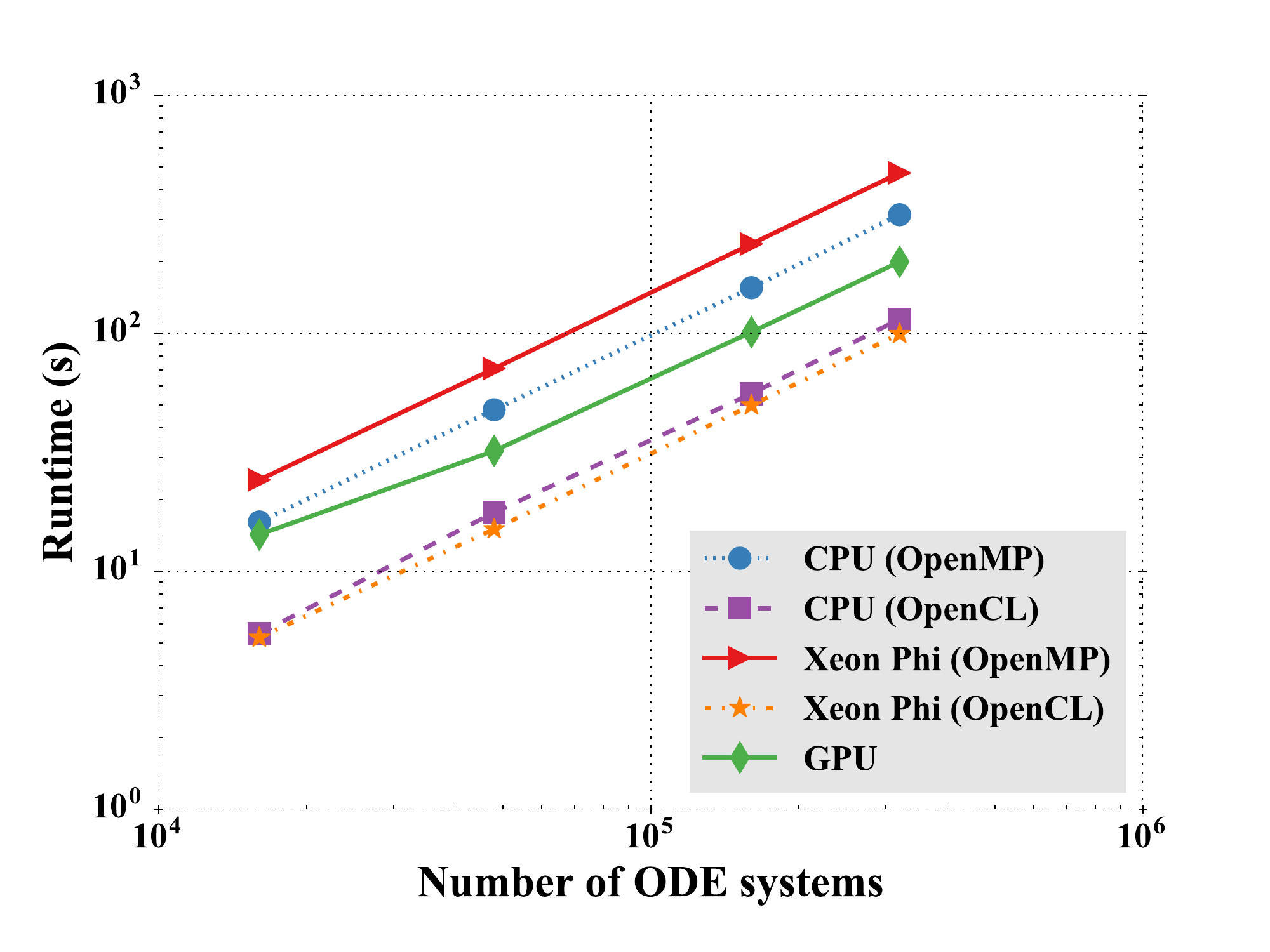}
  \caption{Comparison of runtimes of the RKF45 solver for the model problem, between OpenMP and CL-SIMD on the host CPU and Xeon Phi, and the GPU.
  Data, plotting scripts, and figure file are available~\cite{figures}.}
  \label{fig:rk_runtime}
\end{figure}

\begin{figure}[htbp]
  \centering
  \includegraphics[width=0.9\textwidth]{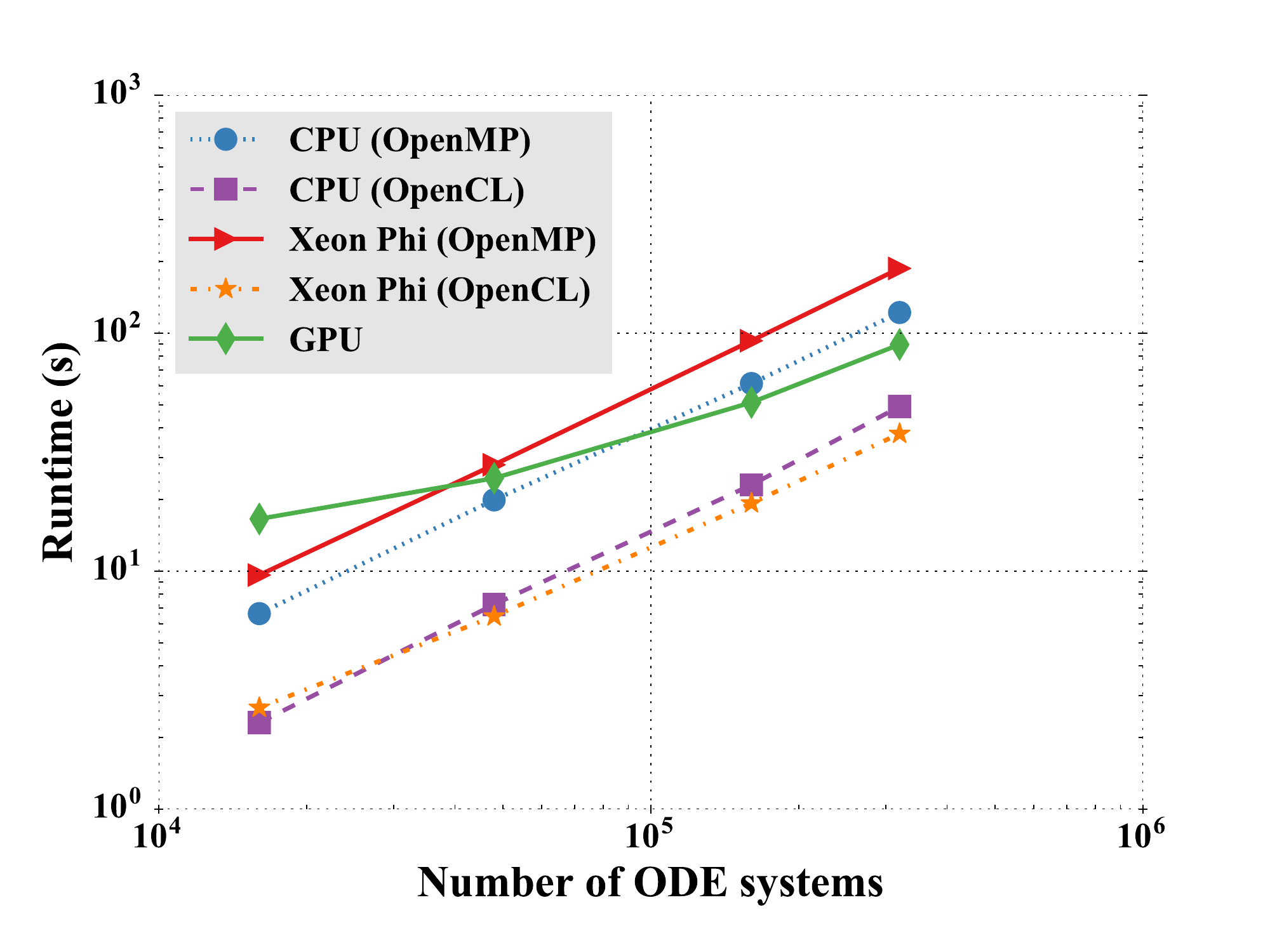}
  \caption{Comparison of runtimes of the ROS4 solver for the model problem, between OpenMP and CL-SIMD on the host CPU and Xeon Phi, and the GPU.
  Data, plotting scripts, and figure file are available~\cite{figures}.}
  \label{fig:ros_runtime}
\end{figure}


\begin{figure}[htbp]
  \centering
  \includegraphics[width=0.9\textwidth]{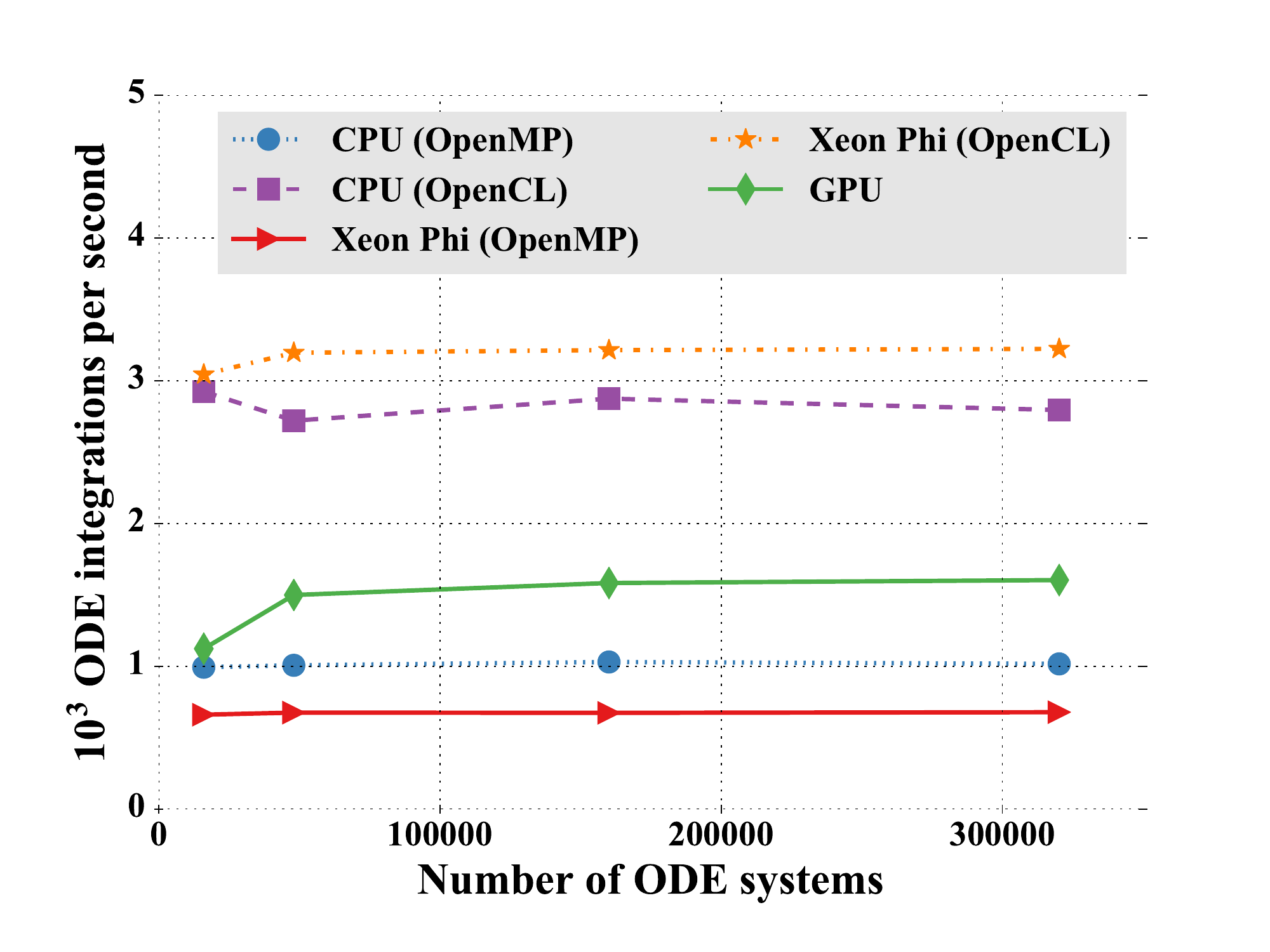}
  \caption{Comparison of computational throughput, measured in \num{e3} ODE integrations per second, of the RKF45 solver for the model problem, between OpenMP and CL-SIMD on the host CPU and Xeon Phi, and the GPU.
  Data, plotting scripts, and figure file are available~\cite{figures}.}
  \label{fig:rk_throughput}
\end{figure}

\begin{figure}[htbp]
  \centering
  \includegraphics[width=0.9\textwidth]{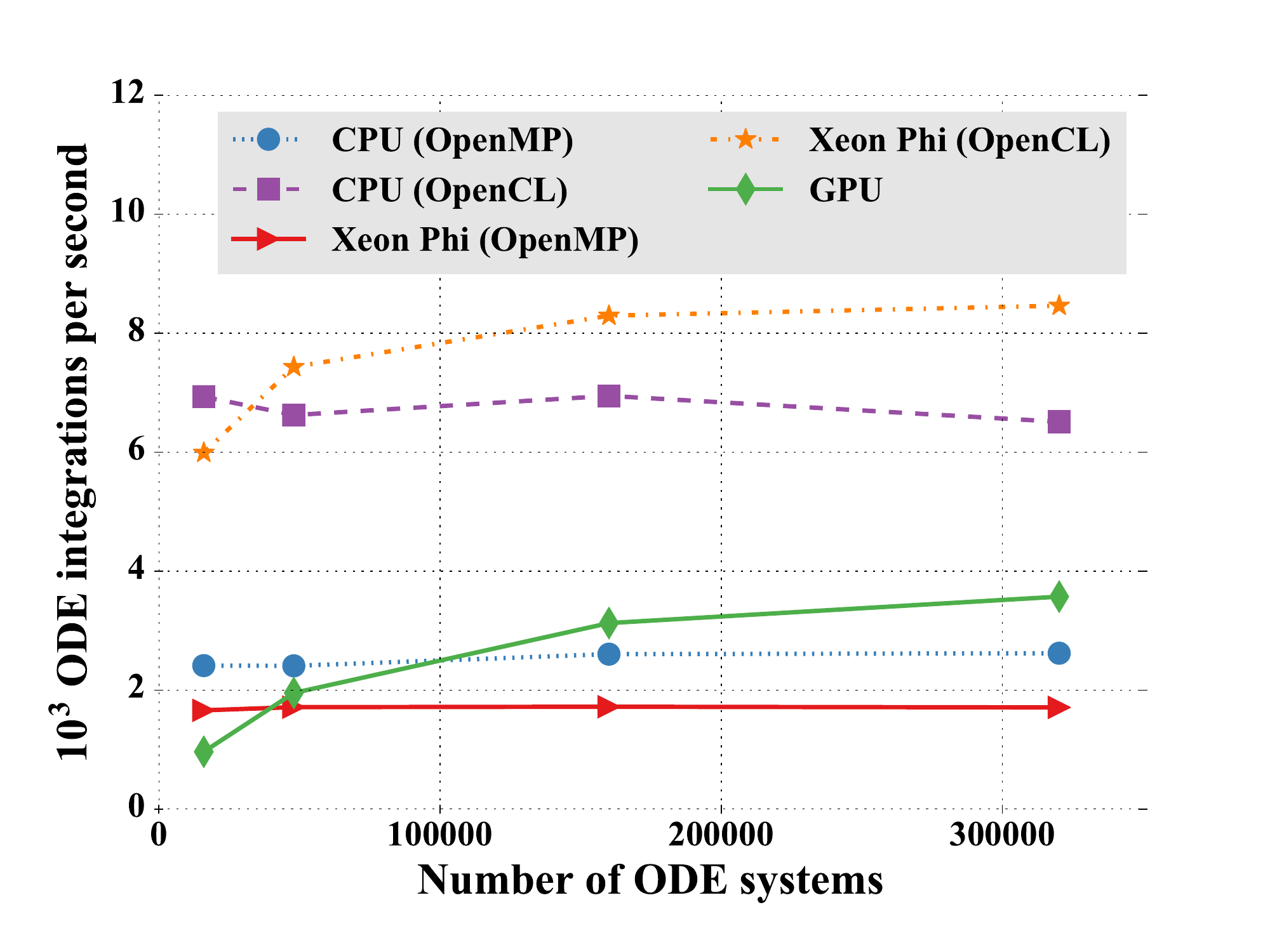}
  \caption{Comparison of computational throughput, measured in \num{e3} ODE integrations per second, of the ROS4 solver for the model problem, between OpenMP and CL-SIMD on the host CPU and Xeon Phi, and the GPU.}
  Data, plotting scripts, and figure file are available~\cite{figures}.
  \label{fig:ros_throughput}
\end{figure}

Figures~\ref{fig:rk_runtime} and \ref{fig:ros_runtime} show the RKF45 and ROS4 runtimes, respectively, for the OpenMP and CL-SIMD methods on the host and MIC accelerator for the  model problem sizes.
The GPU runtimes with CL-thread are also shown for comparison.
The fastest SIMD runtimes are shown based on the word size.
The RHS runtimes all used twice the native word size.
Here, the fastest host performance was found with 16-wide SIMD words (i.e., four times the native word size) while the MIC was fastest using the native size, eight-wide.

The high cost of the ODE integration with RKF45 is evident even for the smallest problem size.
There, tens of seconds are needed for the reaction integrations on the host using the baseline OpenMP method.
(This cost is incurred at least once per global CFD time-step and many thousands of steps may be needed.)
The ROS4 solver is more efficient and consistently \SI{2.5}{$\times$} faster on the host on the model problem.
The ROS4 to RKF45 runtime ratio is identical on the MIC accelerator using OpenMP.

The runtimes scale linearly with the number of ODE systems solved on the host and MIC accelerator using OpenMP.
This is not unexpected as the number of ODE systems solved concurrently on these devices using OpenMP is small relative to the total problem size; that is, the parallelism is small relative to the total problem size.

The ODE systems require different numbers of iterations (see Figure~\ref{fig:flame_profile}) and this leads to variability in the computational cost.
Dynamic loop scheduling, with granularity of one, was used with OpenMP to account for the variable costs.
A strategy was implemented in all OpenCL versions of the integrators to mimic this type of dynamic scheduling.
Here, a simple queue was created using a global counter incremented atomically (i.e., lock-free) by each parallel instruction stream.
Instead of fetching one ODE from the queue as in the OpenMP implementation, the OpenCL version fetches the SIMD (or SIMT) parallel width, i.e., \numlist{4;8;16;32} ODEs depending upon the platform.
This leads to coarse-grained dynamic scheduling and does not address variable costs within each parallel stream.
This impact will be discussed subsequently.

The relative performance differences between the devices and data parallelism methods are more clearly seen by examining the throughput instead of runtime.
The throughputs, defined as the number of ODE systems solved per second, for the RKF45 and ROS4 methods are shown in Figures~\ref{fig:rk_throughput} and \ref{fig:ros_throughput}, respectively.
For the RKF45 method, the throughput is nearly constant for \num{48000} ODE systems and higher.
The SIMD method on the host gives a speedup of \SIrange{2.7}{2.9}{$\times$} over the baseline host OpenMP run-time.
On the MIC accelerator, the speedup is up to \SI{4.8}{$\times$} over OpenMP on the MIC.
That is, the SIMD speedup on the MIC is approximately double that observed on the host, which matches the ratio of the native SIMD word widths on the two devices.

Figure~\ref{fig:ros_throughput} clearly shows the superior throughput of the ROS method.
A more pronounced dependency upon the number of ODE systems is observed, particularly with the MIC SIMD method and the GPU method.
This is driven by the thousands of ODE systems needed to saturate the device with both of these methods, while only tens or hundreds are needed with the other methods.
The ratio of the RKF45 and ROS peak throughputs differs across the platforms: the lowest at 2.2 with the GPU, and the highest of 2.6 with the host and MIC SIMD.

Of note is the lower GPU performance compared with that observed for the RHS function evaluations.
The maximum GPU throughputs are only \SI{1.6}{$\times$} (RKF45) and \SI{1.4}{$\times$} (ROS4) higher than the baseline host OpenMP throughput, yet throughput was \SI{2.3}{$\times$} higher for the RHS function evaluations with GRI Mech 3.0 (see Figure~\ref{fig:rhs_GRI Mech 3.0_best_speedup}).
The host SIMD methods are also lower but to a lesser extent, i.e., 2.7 and \SI{2.5}{$\times$} compared to 3.1 seen previously with the RHS evaluations.
Conversely, the speedup with the MIC SIMD methods (\SI{3.2}{$\times$} for both) are higher relative to the RHS function evaluation benchmark (\SI{2.3}{$\times$}).

A possible cause of this lower performance on the GPU is variability in the number of integrator iterations needed between neighboring ODE systems.
As noted above, there is significant variability in the number of RKF45 iterations between neighbor mesh points.
A unique GPU thread solves each ODE system, which means that the realized cost will be the maximum number of iterations needed by any thread within the same \emph{thread warp}.
The ODE systems in Figure~\ref{fig:flame_profile} are mapped linearly to the GPU threads within each warp.
Recall that the only variability between RHS function evaluations was the temperature polynomials, a relatively minor cost.
Performance degradation caused by differing numbers of integration iterations has been reported before by Stone and Davis~\cite{Stone:2013b} and Niemeyer and Sung~\cite{Niemeyer:2014jcp}.

Problem-to-problem variability should also impact the SIMD implementations.
We define an inefficiency metric to better assess this performance impact of the ODE variability: the \emph{waste} within each SIMD work unit (i.e., a SIMD word or SIMT warp) represents the number of excess integrator steps taken.
The cost per integrator step is constant for the RKF45 and ROS4 methods so this is a logical quantity to measure.
The waste within SIMD work unit $j$ can be expressed in a normalized form as
\begin{equation}
W_j = 1 - \frac{\sum_i^{P} N_i}{P \times \max_i^{P} N_i} \;,
\label{eqn:waste}
\end{equation}
where $N_i$ is the number of integrator iterations needed for ODE $i$ and $P$ is the parallel width (i.e., 4, 8, 16, or 32 depending upon the device).
Equation~\eqref{eqn:waste} extends the CUDA-specific \emph{warp} divergence metric proposed by Niemeyer and Sung~\cite{Niemeyer:2014jcp} to any SIMD (or SIMT) platform with vector length $P$.

\begin{figure}[htbp]
  \centering
  \includegraphics[width=0.8\textwidth]{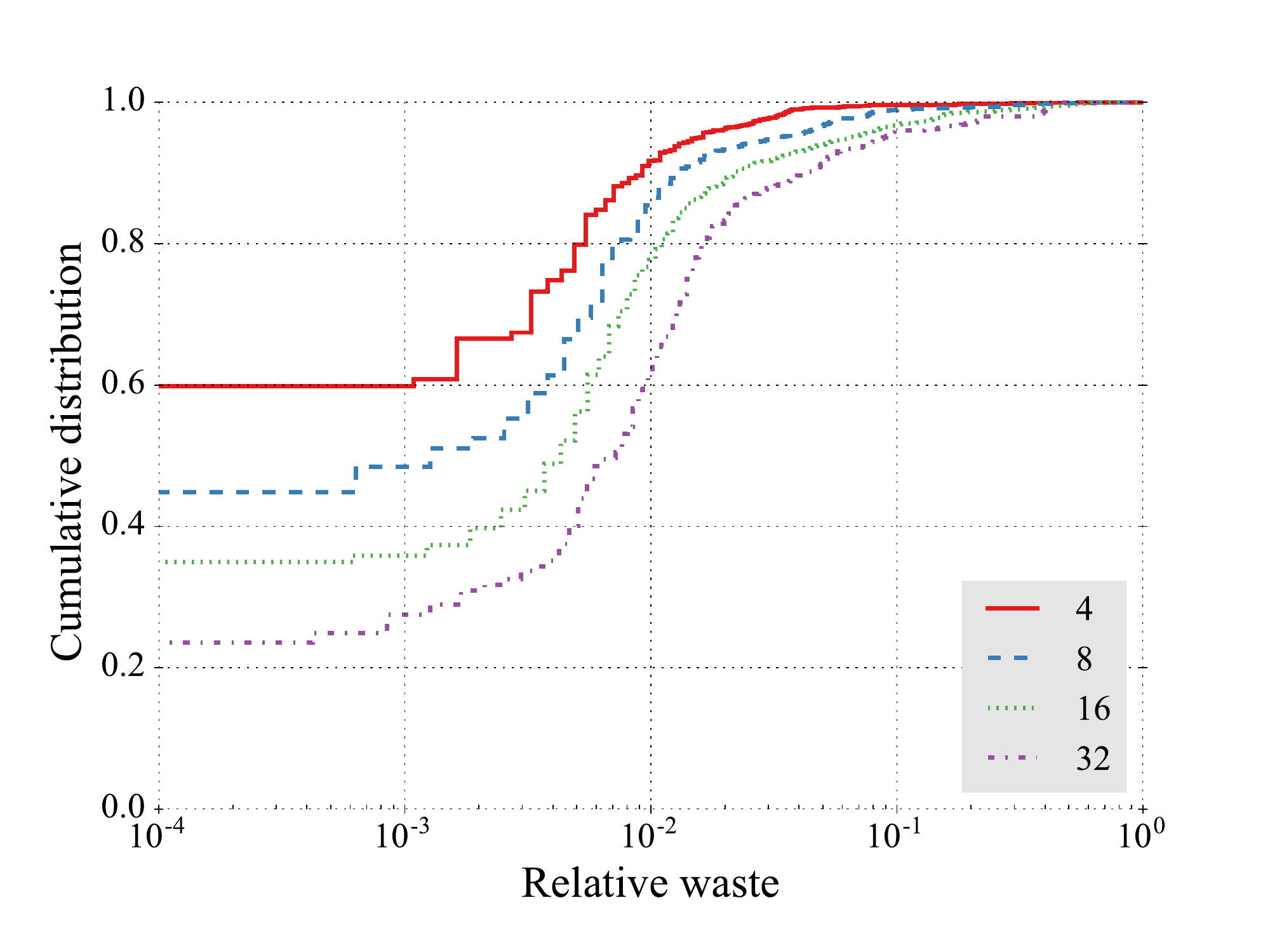}
  \caption{Cumulative population of relative waste in parallel SIMD work, given by Eq.~\eqref{eqn:waste}, for the RKF45 solver on the model problem with SIMD parallel widths 4, 8, 16, and 32. Relative waste calculated from sample of 1601 points.
  Data, plotting scripts, and figure file are available~\cite{figures}.}
  \label{fig:rk_waste}
\end{figure}

\begin{figure}[htbp]
  \centering
  \includegraphics[width=0.8\textwidth]{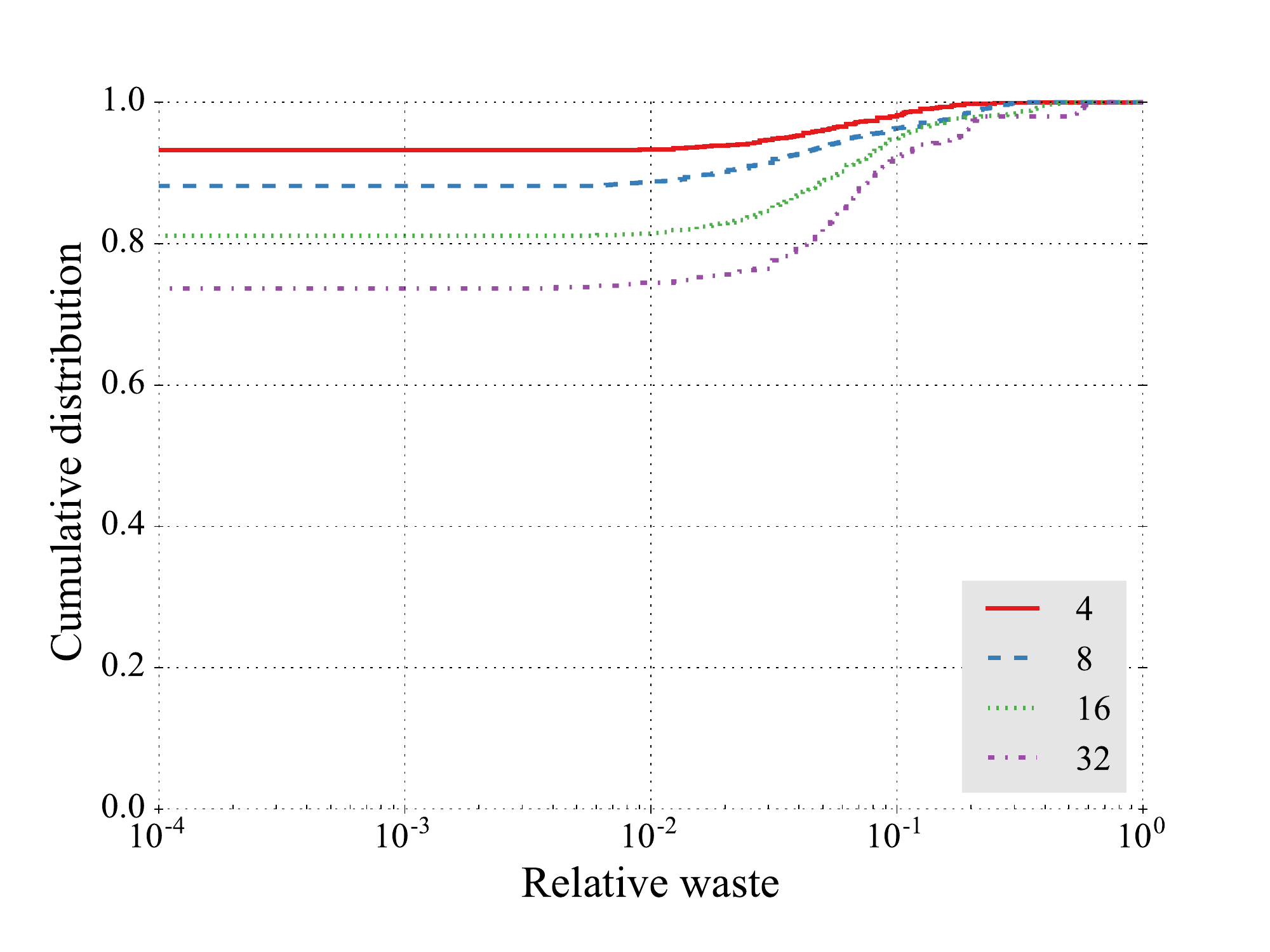}
  \caption{Cumulative population of relative waste in parallel SIMD work, given by Eq.~\eqref{eqn:waste}, for the ROS4 solver on the model problem with SIMD parallel widths 4, 8, 16, and 32. Relative waste calculated from sample of 1601 points.
  Data, plotting scripts, and figure file are available~\cite{figures}.}
  \label{fig:ros_waste}
\end{figure}

Figures~\ref{fig:rk_waste} and \ref{fig:ros_waste} show the cumulative probability distribution of $W$ for the model problem for the four relevant SIMD widths for RKF45 and ROS4, respectively.
The impact of wider SIMD parallelism is evident for both solvers.
That is, as the SIMD width is increased, the proportion of wasted computation increases.
For RKF45, we see that approximately \SI{92}{\percent} of the work units have less than \SI{1}{\percent} waste with a width of four, but this drops to only \SI{63}{\percent} for a width of 32.
We observe a similar behavior with ROS4, though with reduced magnitudes.
This is consistent with Figure~\ref{fig:flame_profile} where the variation between adjacent ODE systems (i.e., mesh points) is less.

\begin{figure}[htbp]
  \centering
  \begin{subfigure}{0.7\textwidth}
      \includegraphics[width=\linewidth]{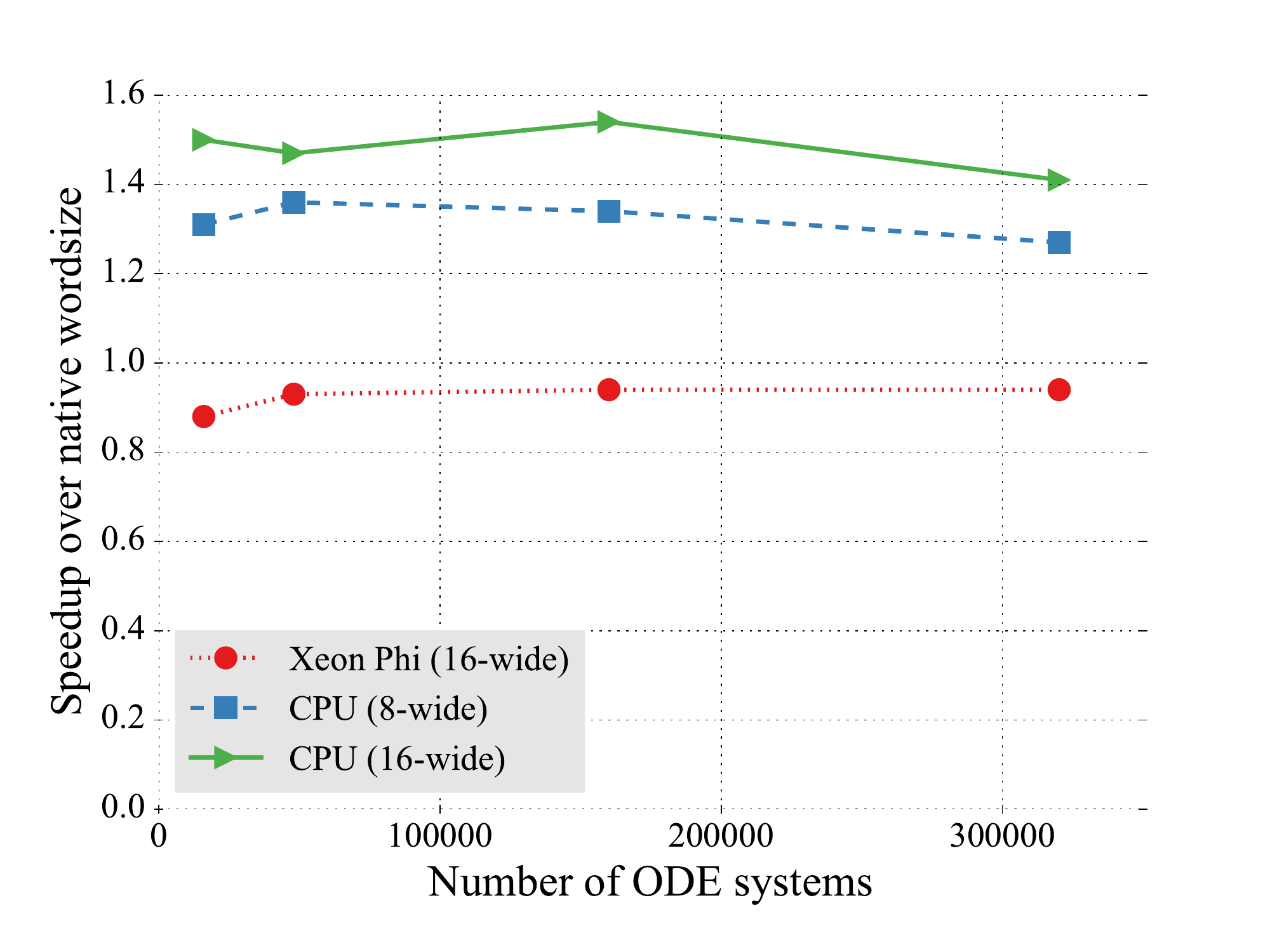}
      \caption{RKF45}
  \end{subfigure}
  \\
  \begin{subfigure}{0.7\textwidth}
      \includegraphics[width=\linewidth]{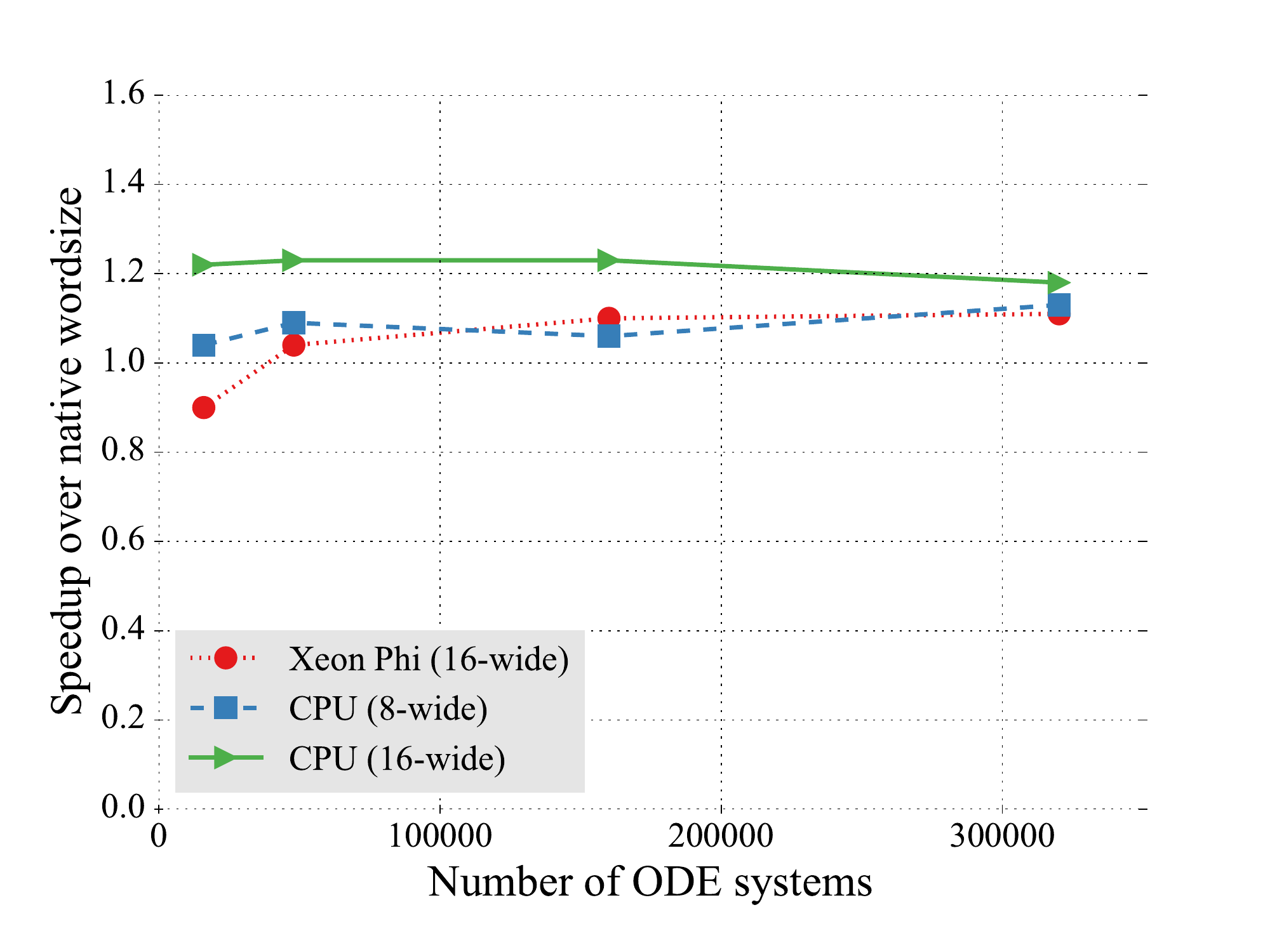}
      \caption{ROS4}
  \end{subfigure}
  \caption{Speedup of integrators using SIMD word sizes of 16-wide on the Xeon Phi, and 8- and 16-wide on the CPU, which represent two, two, and four times the native word sizes, respectively.
  Data, plotting scripts, and figure file are available~\cite{figures}.}
  \label{fig:speedup_wordsize}
\end{figure}

Figure~\ref{fig:speedup_wordsize} shows the relative throughput of the SIMD versions of the ODE solvers on the host and MIC using increasing SIMD word sizes.
There, the throughput on each device is normalized by the throughput with the native SIMD word size.
Specifying an SIMD word larger than the native size (e.g., \texttt{double8} on the host) should result in multiple SIMD operations in sequence.
The analysis above predicts that the performance, especially with the RKF45 solver, should degrade with wider word size.
However, using wider words significantly improves performance on the host.
In fact, the highest performance on the host with both RKF45 and ROS4 is obtained using \texttt{double16}, four times the native word size, and the performance consistently improves using wider SIMD words.
On the MIC, the wider word size degrades performance of the RKF45 solver by approximately \SI{10}{\percent}.
On the other hand, the wider word size using the ROS4 solver improves the MIC performance by approximately \SI{10}{\percent}.
The performance metrics presented here indicate that while problem-to-problem variation leads to increased computational waste, this does not necessarily translate into reduced computational throughput on the host and MIC devices.

\section{Conclusions}

In this paper, we presented and discussed the parallel performance of thread- and data-parallel methods applied to chemical kinetics integrations.
Benchmarks were conducted using multithreading and SIMD parallel methods on a multicore CPU system and on two coprocessors (or accelerators): an Intel Xeon Phi (or MIC) and an Nvidia Kepler K20m GPU.
We implemented both multithreading and data-parallel models using OpenCL to allow a study of the same code base across all three platforms.
Two vectorization models were examined within OpenCL: (1) one thread maps to each parallel task and sets of cooperating threads execute instructions in a SIMT paradigm, and (2) the same instruction stream concurrently computes explicit SIMD vector datatypes and multiple parallel tasks.

All benchmarks were compared with multicore runs on the host CPU and the MIC using OpenMP.
We were particularly interested in the performance difference between OpenMP with automated compiler vectorization compared with manual SIMD programming on the host and MIC platforms.

The first benchmark series studied the performance of evaluating many instances of the RHS function for three chemical kinetic models of increasing size and complexity.
The SIMT implementations on the host and MIC devices both perform slower (\SI{13}{\percent} and \SI{35}{\percent}, respectively) than their baseline OpenMP implementations with the common GRI Mech 3.0 model~\cite{grimech3}.
However, the explicit SIMD model provides a speedup of approximately \SI{3}{$\times$} over the baseline OpenMP on both of these platforms.
The SIMT model on the GPU performs far better than the SIMT model on the host and MIC devices, and matches the performance of the MIC device with SIMD programming.
Thus, while both SIMD and SIMT models are possible on CPU and the CPU-like MIC using OpenCL, SIMD methods provide considerably higher performance.
Furthermore, SIMT is necessary on the GPU platform meaning that two separate programming models are needed to reach peak performance across the three HPC devices.

Studies with all three chemical kinetic models produced similar results, and performance showed no significant dependence upon model size.
For this reason, we used only the GRI Mech 3.0 model for subsequent benchmarks.

The second benchmark series studied the performance of integrating many independent constant-pressure ODE systems with the GRI Mech 3.0 model for methane oxidation.
This model problem mimicked what is commonly encountered when simulating chemical kinetics phenomena within an operator-splitting framework.
The ODE systems were integrated using the nonstiff RKF45 ODE solver and stiff Rosenbrock ROS4 solver.
Both methods have the same fourth-order theoretical accuracy and use the same step-size adapation and initial step-size estimation algorithms.

The ROS4 solver consistently performs \numrange{2.2}{2.6} times faster than the RKF45 solver on the model problem on the various platforms.
This matches analysis that shows RKF45 needs approximately 25 times more iterations than ROS4, while ROS4 costs approximately ten times more than RKF45 per iteration (based on the ratio of RHS function evaluations).
Finite-difference Jacobian matrices were used with ROS4 for this study for simplicity, which increased the number of RHS evaluations per-iteration by $\NSpecies+1$---an increase of nine times for GRI Mech 3.0.
Analytical Jacobian matrices for the model constant-pressure problem have been derived \cite{Safta:2011,Niemeyer:2016aa} and can reduce the cost per-iteration of the Rosenbrock family of solvers.
(Stone and Bisetti~\cite{Stone:2014a} showed a 2.9 times speedup with analytical Jacobian matrices for GRI Mech 3.0 with ROS4.)

The SIMD implementation of the two solvers shows a significant performance acceleration compared with the baseline OpenMP implementation.
In general, the SIMD method improves performance by \numrange{2.5}{2.8} times on the host and \numrange{4.7}{4.9} times on the MIC.
The higher MIC SIMD acceleration is consistent with the ratios of the SIMD word widths.

The GPU ODE integrators do not perform as efficiently as the RHS function evaluation.
The GPU integrator only offers a \numrange{1.4}{1.6} times speedup over the host baseline throughput.
We attributed the lower performance to thread divergence caused by ODE systems requiring different numbers of integrator steps in each SIMT parallel work unit.
We quantified this impact with a SIMD waste metric that shows that the wasted number of integrator steps increases with increasing vector width (32 for the GPU).
The ROS4 integrator exhibits a lower occurrence of wasted work, which can help explain why the ROS integrator performed better than the RKF45 integrator on the GPU and MIC devices.
However, the performance on the host and MIC often \emph{improves} with increasing SIMD word size.
This indicates that the improved computational performance due to wider word sizes (e.g., instruction parallelism, cache efficiency) can actually overcome increased integrator waste.
Nevertheless, reducing the wasted number of integrator steps could improve the performance on all devices and should be investigated in future studies.
For example, Murray~\cite{Murray:2012} demonstrated a strategy of mitigating inefficiency due to variable-length RK integration tasks (i.e., variation in steps) on GPUs by assigning multiple tasks to each GPU thread.
This strategy may be extendable to SIMD platforms and should be investigated in future studies to determine the impact of variable task lengths.

Overall, this paper shows that the explicit SIMD methods offer a promising strategy for more efficiently using MIC and host CPU systems within the context of chemical kinetics applications.
Further research is needed to improve the SIMD-friendly ROS methods, e.g., with analytical Jacobian matrices.
The SIMD performance advantages demonstrated here may warrant investigation into more numerically efficient, but less SIMD efficient, ODE solver methods such as implicit Runge--Kutta integrators.

\section*{Acknowledgements}

This material is based upon work supported, in part, by the National Science Foundation under grant ACI-1535065.
This work used the Extreme Science and Engineering Discovery Environment (XSEDE), which is supported by National Science Foundation grant number ACI-1053575.
Code development and performance measurements were conducted on the Stampede system at the Texas Advanced Computer Center (TACC) through resource allocation TG-ASC130025.

\appendix

\section{Availability of material}

The integrators used to perform this study are available openly via the \texttt{accelerInt}
software package~\cite{accelerInt}.
The most recent version of \texttt{accelerInt} can be found at its GitHub repository:
\url{https://github.com/SLACKHA/accelerInt}.
All figures, and the data and plotting scripts necessary to reproduce them, are available openly under the CC-BY license~\cite{figures}.

\section{RKF parameters}
\label{app:rkf}

Table~\ref{tab:rkf45} shows the method parameters in a modified Butcher tableau, where $a_{ij}$ are the coefficients, $b_j$\slash $\hat{b}_j$ are the weights of the embedded fourth-order and fifth-order methods, respectively, and $c_i$ are the nodes (not used here, since the ODE systems are autonomous).

\begin{table}[htbp]
\centering
\begin{tabular}{@{}c c c c c c c c@{}}
\toprule
$c_i$ & \multicolumn{5}{c}{$a_{ij}$} \\
\cmidrule(r){1-1} \cmidrule(lr){2-6}
0 \\
$\frac{1}{4}$   & $\frac{1}{4}$ \\
$\frac{3}{8}$   & $\frac{3}{32}$      & $\frac{9}{32}$ \\
$\frac{12}{13}$ & $\frac{1932}{2197}$ & $-\frac{7200}{2197}$ & $\frac{7296}{2197}$ \\
1               & $\frac{439}{216}$   & $-8$                 & $\frac{3680}{513}$   & $-\frac{845}{4104}$ \\
$\frac{1}{2}$   & $-\frac{8}{27}$     & 2                    & $-\frac{3544}{2565}$ & $\frac{1859}{4104}$ & $-\frac{11}{40}$ \\
\midrule
$b_j$           & $\frac{25}{216}$    & 0                    & $\frac{1408}{2565}$  & $\frac{2197}{4104}$ & $-\frac{1}{5}$  & 0 \\
\midrule
$\hat{b}_j$     & $\frac{16}{135}$    & 0                    & $\frac{6656}{12825}$  & $\frac{28561}{56430}$ & $-\frac{9}{50}$  & $\frac{2}{55}$ \\
\bottomrule
\end{tabular}
\caption{Coefficients for the five-stage, fourth-order Runge--Kutta--Fehlberg method, adopted from Hairer et al.~\cite{Hairer:1993} and displayed in a modified Butcher tableau.}
\label{tab:rkf45}
\end{table}

\section{ROS4 parameters}
\label{app:ros4}

Table~\ref{tab:ros4} contains the ROS4 parameters, including the strictly lower-triangular matrices $\myMatrix{a}$ and $\myMatrix{c}$, and the vectors $\myVector{m}$, $\myVector{b}$, $\alpha_i$, and $\gamma_i$.
In addition, it shows the vector $\myVector{E}$, the difference in $\myVector{b}$ coefficients for method orders three and four used for error estimation.

\begin{table}[htbp]
    \centering
\begin{tabular}{@{}l l@{}}
\toprule
$\begin{aligned}[t]
    a_{21} &=  2.0 \\
    a_{31} &=  1.867943637803922 \\
    a_{32} &=  0.2344449711399156 \\
    a_{41} &=  a_{31} \\
    a_{42} &=  a_{32} \\
    c_{21} &= -7.137615036412310 \\
    c_{31} &=  2.580708087951457 \\
    c_{32} &= 0.6515950076447975 \\
    c_{41} &= -2.137148994382534 \\
    c_{42} &= -0.3214669691237626 \\
    c_{43} &= -0.6949742501781779 \\
    b_{1} &= 2.255570073418735 \\
    b_{2} &= 0.2870493262186792
\end{aligned}$ &
$\begin{aligned}[t]
    \alpha_{2} &= 1.14564 \\
    \alpha_{3} &= 0.6552168638155900 \\
    \alpha_{4} &= \alpha_{3} \\
    \gamma_{1} &= \gamma = 0.57282 \\
    \gamma_{2} &= -1.769193891319233 \\
    \gamma_{3} &= 0.7592633437920482 \\
    \gamma_{4} &= -0.1049021087100450 \\
    E_{1} &= -0.2815431932141155 \\
    E_{2} &= -0.07276199124938920 \\
    E_{3} &= -0.1082196201495311 \\
    E_{4} &= -1.093502252409163 \\
    b_{3} &= 0.4353179431840180 \\
    b_{4} &= 1.093502252409163
\end{aligned}$ \\
\bottomrule
\end{tabular}
    \caption{Parameters for ROS4, adopted from Hairer and Wanner~\cite{ROS4:1992}.}
    \label{tab:ros4}
\end{table}

\bibliographystyle{elsarticle-num}
\bibliography{refs}

\end{document}